\documentstyle[12pt]{article}
\title{A model for $q^2\bar{q}^2$ systems, \\
illustrated by an application to $K\bar{K}$ scattering}
\author{B.Masud\thanks{E-mail address
(Internet):MASUD@PHCU.HELSINKI.FI},\\
 Research Institute for Theoretical Physics,  \\ P.O.Box 9,
SF-00014, University of Helsinki, Finland.}

\begin{document}
\setlength{\baselineskip}{5ex}

\maketitle

\begin{abstract}
Here is  presented a four-body potential model for $q^2\bar{q}^2$
systems which includes both
the  spin and flavour degrees of freedom, extending the
formalism presented already in
the spin independent situation. This allows an application
to a realistic situation, which is chosen to be
 $K\bar{K}$ scattering. It is seen that
because of  the gluonic effects   in this multi-quark system, the
$K\bar{K}$
attraction resulting from the quark-exchange mechanism gets
appreciably
decreased compared to that emerging through
the naive two-body potential approach.

%\vspace{2.5ex}
% The assignement of the PCAS nos. is according to the AIP  classification
%scheme.
\noindent PCAS number(s):12.40.Qq, 13.75.Lb,  12.38.Lg, 14.40.Cs

\end{abstract}

\renewcommand{\theequation}{\thesection.\arabic{equation}}
\renewcommand{\thetable}{\thesection.\arabic{table}}
\section{Introduction}
\setcounter{equation}{0}
With strong evidence in favour of quark
confinement both from experiment (failure to find free quarks) and
theory  \cite{band81,kogu83}, we have some understanding of the
quark-quark
interaction
for large distances. Using this, along with the understanding of
the short distance quark-quark interaction obtained through
perturbative QCD, different models of
the quark-quark interaction  have been used such as the MIT bag
model~  \cite{chod974,chod1074} and the
constituent quark potential model. In the
constituent quark potential model the quark-antiquark
interaction is represented by a
potential which is well motivated by QCD for small distances
(given by
the one-gluon exchange mechanism), but is
modified so as to incorporate the confining potential as the
limit for large
distances. Moreover, the current masses of the QCD lagrangian
are replaced by
effective masses, termed {\em constituent masses}, which
are fitted, along
with other parameters of  the model, to  experimentally known
quantities
related to some set of hadrons. For quarks  of known (fitted) masses
interacting through a space dependent potential, one can set up and
solve a
Schr\"{o}dinger equation for their dynamics.
In this way the constituent quark potential model, when improved
to incorporate
relativistic effects,
 explains in a consistent and unified way most of the observed
mesonic states, from the pion to the upsilon,  as quark-antiquark
states with
different values of orbital and radial quantum
numbers~\cite{godi85}.

But that is not enough.
A successful model of strong interactions should be able to
describe also
possible systems having three or more quarks/antiquarks.
It is not clear yet how, if at all, the quark potential model
can be applied to multi-quark systems (with more than three
quarks).
Perhaps the simplest approach is to take the
many-body hamiltonian as a sum of hamiltonians corresponding
to all pairs of
particles involved, the basic method used in
\cite{wn83,wn90,barn92,barn93}.
But this has many theoretical as well as phenomenological
(such as the van der
Waals force problem   \cite{greb81}) flaws. Keeping this in mind,
a {\em four-body} potential model for
a quark-exchange mechanism in $q^2\bar{q}^2$ systems
was proposed in
\cite{masu91} in a spin independent situation,
taking  into account the
effects of the gluonic degrees of freedom as well in a
non-trivial way.
Only for small distances does this agree with
the sum of two-body (${\bf F}_i\cdot{\bf F}_j$) potentials model.
For, roughly speaking, interquark distances greater than 0.5 fm it
qualitatively agrees with the flux tube model   \cite{pis85,merl87}
of the
adiabatic surfaces of the gluonic field.

As shown in   \cite{pis85},
there are three flux tube topologies for the $N_q=N_{\bar{q}}=2$
system, each of which will determine the ground state in a
different region of configuration space.
Since these are linearly independent of each other, we need a basis
containing
at least {\em three} gluonic states to describe the gluonic field of
this multiquark system.  This is different to what the two-body
potential model
says for the $q^2\bar{q}^2$ system; the two-body model
would need only a {\em
two} dimensional colour basis~  \cite{wn90}.
Thus in   \cite{masu91} the two-body potential model was written
in a redundant colour basis
\begin{equation}
\label{d123}
{|1\rangle }_c={|{\bf 1}_{1\bar{3}}{\bf 1}_{2\bar{4}}\rangle }_c,
{|2\rangle }_c= {|{\bf
1}_{1\bar{4}}{\bf 1}_{2\bar{3}}\rangle }_c {\rm \ \ \ and \ \ \ }
{|3\rangle }_c={|{\bf
\bar{3}}_{12}{\bf 3}_{\bar{3}\bar{4}}\rangle }_c,
\end{equation}
corresponding to the three
basic states in the flux tube model, and then the
suggested changes, forming
the proposed model, were made. The (confinement)
potential part of the hamiltonian of the
system was written as
\begin{eqnarray}
\label{nvform}
V(q_1q_2\bar{q}_3\bar{q}_4) & =  &
\sum_{i<j}{\bf F}_i\cdot{\bf F}_jv_{ij}, \ \ \ \ \
 \ \ \ \ \ {\rm with} \\
\label{nqd}
v_{ij} & = &  Cr_{ij}^2+\bar{C}.
\end{eqnarray}
This quadratic, rather than the theoretically better motivated
coulomb-plus-linear, form of the two-body potential was used
basically for
computational convenience. Similarly
for the kinetic energy part, the non-relativistic
expression was used, also for simplicity and numerical convenience.
These features have been retained in the present work as well.

The overlap matrix $N$ in this redundant basis is as
given by eq.(2.8) of
  \cite{masu91}, the potential matrix by eq.(3.10) there,
and the matrix element of the non-relativistic
kinetic energy operator between any two
states is expressed through eq.(3.11)
of the same (these are also
given by eqs.(\ref{nsom}),(\ref{nvfom}) and
(\ref{nkom}) below respectively, with $f=1$).  The model proposed in
  \cite{masu91} gives the following expressions for these matrices:
\begin{eqnarray}
\label{nsom}
N\rightarrow N(f) & = & \left( \begin{array}{ccc} 1 & \frac{1}{3}f &
\sqrt{\frac{1}{3}}f \\ \frac{1}{3}f & 1 & -\sqrt{\frac{1}{3}}f \\
\sqrt{\frac{1}{3}}f & -\sqrt{\frac{1}{3}}f &
1 \end{array} \right), \\
_g\langle X'|H|X\rangle _g & = &
\ _g\langle X'|K|X\rangle_g+
\ _g\langle X'|V|X\rangle _g, \nonumber \\
\label{nkom}
_g\langle X'|K|X\rangle _g & = & N(f)_{X',X}^{1/2}
\left(\sum_i-\frac{\underline{\nabla}_i^2}{2m_i}
\right)N(f)_{X',X}^{1/2}  \ \ \ \ \ {\rm and} \ \ \ \ \
\end{eqnarray}
\begin{equation}
\label{nvfom}
V= \left( \begin{array}{ccl}
-\frac{4}{3}(v_{1\bar{3}}+v_{2\bar{4}}) &
\frac{4f}{9}\left( \begin{array}{c} v_{12}+v_{\bar{3}\bar{4}}
\\ -v_{1\bar{3}}-v_{2\bar{4}} \\ -v_{1\bar{4}}-v_{2\bar{3}}
\end{array} \right) &
\frac{2f}{3\sqrt{3}}\left( \begin{array}{c}
-2(v_{1\bar{3}}+v_{2\bar{4}}) \\ +v_{1\bar{4}}
+v_{2\bar{3}} \\ -v_{12}-v_{\bar{3}\bar{4}} \end{array} \right) \\  &
-\frac{4}{3}(v_{1\bar{4}}+v_{2\bar{3}}) & \frac{2f}{3\sqrt{3}}\left(
\begin{array}{c} 2(v_{1\bar{4}}+v_{2\bar{3}}) \\ +v_{12}
+v_{\bar{3}\bar{4}} \\
-v_{2\bar{4}}-v_{1\bar{3}} \end{array} \right) \\
{\rm symmetric} &  & \begin{array}{l}
-\frac{1}{3}\left( \begin{array}{c}
2(v_{12}+v_{\bar{3}\bar{4}}) \\ +v_{1\bar{3}}
+v_{2\bar{4}} \\ +v_{1\bar{4}}+v_{2\bar{3}} \end{array}
\right) \\ +\frac{5}{2}Df(1-f) \end{array}
\end{array} \right).
\end{equation}

The basis now is actually $|1\rangle _g,|2\rangle _g \
{\rm and} \ |3\rangle _g$, instead of   $|1\rangle _c,|2\rangle _c \
{\rm and} \ |3\rangle _c$; the new
subscript {\em g} refers to the
gluonic degree of freedom, instead of the
colour degree of freedom represented by the subscript {\em c}.
$H$ is the hamiltonian of the system, and
$N(f)_{X',X}$ is the gluonic states overlap factor
$_g\langle X'|X\rangle _g$.
The matrices in the above three equations
are achieved by multiplying the two-body potential
model expressions for the off-diagonal
elements of overlap, kinetic energy and
potential energy matrices respectively by a factor $f$ which
tends to 1 for all $r_{ij}  \ll   b_s^{-1/2}$ and to 0 when any
$r_{ij}$ becomes $\gg  b_s^{-1/2}$.
In the proposed model in   \cite{masu91},
the diagonal elements were the same  as
obtained in the two-body model calculations, apart from
the $\frac{5}{2}Df(1-f)$ term in the 3,3 element of the proposed
potential matrix (eq.(\ref{nvfom})).
The reasons for the introduction of this term
are described in the discussion from eq.(4.18) till
the end of section 4 of   \cite{masu91}.
It should be clear from the
discussion there that the model is actually defined
in a basis which is {\em not}
redundant even for all $r_{ij}$ vanishing.

This treatment of the diagonal and the
off-diagonal elements of the two-body
model based matrices was
motivated by the work presented in    \cite{gp89,glu89},
along with the similarity
for large interquark distances in space dependence of the  diagonal
elements of the $K, V \ {\rm and} \ N$ matrices in both the two-body
potential and the flux tube model.  The calculations reported in
\cite{gp89,glu89} actually show
that in the flux tube model the coupling of the three gluonic states
${|1\rangle }_g$, ${|2\rangle }_g$ and ${|3\rangle }_g$
decreases exponentially with
inter-quark distances.
For the space dependent factor $f$ multiplying the off-diagonal
elements,
the choice used in   \cite{masu91} was
\begin{equation}
\label{f}
f=\exp (-\bar{k}\sum_{i< j}r_{ij}^2) \ \ \ , \ \ \
\bar{k}=\frac{1}{6}kb_s
\end{equation}
\noindent with $k$ a numerical coefficient. This is the simplest
choice from
a computational point of view.
An alternative choice of $f$ is suggested in
   \cite{mor89,alm90}. Both of these forms have been
studied in   \cite{mogp91,migp93,migps92}, which aim at
extracting the
gluon field overlap factor $f$ from a calculation using
lattice Monte
Carlo techniques. This work is in progress, but so far
neither of these two candidates is preferred by
the calculations.

Developing  a formalism, based on the model proposed above,
describing the dynamics of meson-meson interactions  also
requires taking into account the ``slower motion'', the quark
position dependence of the wave function. This was done, for the
spin-independent case, in sections 5
and 6 of the previous work   \cite{masu91}. There we used
our model to specify a theoretically refined (but non-relativistic)
hamiltonian of the $q^2\bar{q}^2$
system in our gluonic basis, and  solved {\em approximately} for
the wave
function using the `resonating group method' common
in nuclear   \cite{wildt77} and also
recently in particle physics   \cite{mast87}.
The approximation corresponded to
specifying parts of the quark position dependent
wave function before solving the Schr\"{o}dinger
equation for the rest.

In the present work we
introduce the spin and flavour degrees of freedom, and as a
first application apply
the formalism to  a physical meson-meson system, namely $K\bar{K}$.
This is done in section 2.
The solution for the total wave function of the system thus obtained
gives us numerical results
for the corresponding meson-meson phase shifts  for the
elastic as well as
the non-elastic meson-meson scattering,
along with a condition for the existence
of a bound state of the whole system.
These results are reported in section 3. This is followed by
our conclusions.

\section{The model applied to $K\bar{K}$ systems}
\setcounter{equation}{0}

In this section we present the formalism for  a
realistic situation. This  means that along with the gluonic basis
described in   \cite{masu91}, we have to deal with the spin basis
as well.
Moreover, flavour dependence has to be considered. The hamiltonian
of the
system, written now in this spin-gluonic basis,
would also include the hyperfine
term. To  complete the formalism, we  have to incorporate
also the quark
position dependence of the wave function.
With quark contents of the $q^2\bar{q}^2$ system like that of
$K\bar{K}$, the flavour wavefunction can be written generally
as $l\bar{s}
\bar{l}s$, with $l$ standing for a light (up or down)
quark. We label these four particles  $1,\bar{3},\bar{4}$
and 2 respectively.
Thus the pair $(2,\bar{3})$   would be composed of strange
quark and antiquark, and $(1,\bar{4})$ of light
quark and antiquark.  In this way each of the
particles  2 and $\bar{3}$ has a mass  higher  than
of those belonging to the other pair (1 and $\bar{4}$)
by a ratio which is the same as that of the strange
quark  mass $m_s$ to the up (or down) quark mass $m$. This mass
ratio is denoted by $s$ in this paper.
Anti-symmetrization of the total wave function
is not necessary in the present case since
we do  not have any two identical fermions.

For the spin dependent part of the basis we use, as in
the quark model, the states arising through the spins of the
quarks only. So any pair of particles (quark or
antiquark) would  have a total spin of one or zero, thus
forming a spin
triplet or singlet respectively. This also means that the total spin
of the whole $q^2\bar{q}^2$ system  can have
a value of zero, one  or two. Being  interested  in the ground state
of the $J^P=0^+$ sector of the system, we focus on the spin states
with the total spin of the system as zero. This is
consistent because, as
is mentioned later, tensor and spin-orbit forces are
neglected in this work. Hence, the hamiltonian separately conserves
${\bf L}$ and ${\bf S}$ i.e. the total orbital angular momentum and
the total spin of the system. This allows us to restrict our
considerations to the ${\bf S}=0$ sector, as mentioned
above, meaning,
in turn, that we will be dealing only with ${\bf L}=0$ spatial wave
functions.

In every one of the three channels of the previous section
(corresponding to
the three gluonic states $|1\rangle_g,|2\rangle_g$ and
$|3\rangle_g$), our four particles can be grouped into
two mesonic sub-clusters. Each of these
clusters may have a combined spin of zero or one and hence the
$q^2\bar{q}^2$ system  may be composed of
either two spin singlets  or two triplets.  This means
that there may
be {\em two} independent spin channels for each  of the three
gluonic channels above.  Thus, there are  {\em six} independent
states of the system in hand. The corresponding six spin states are
written in the notation of Appendix D of   \cite{wn90}
(see Appendix A of the
present paper for details) as:

\noindent In the first channel (with the gluonic part of the
base state
 as $|1\rangle_g$):
\begin{equation}
\label{sp1}
|1S\rangle_s =  |P_{1\bar{3}}P_{2\bar{4}}\rangle_s
 \ \ \ {\rm and} \ \ \
|1T\rangle_s  =  |{\bf V}_{1\bar{3}}
\cdot{\bf V}_{2\bar{4}}\rangle_s.
\end{equation}
In the second channel:
\begin{equation}
\label{sp2}
|2S\rangle_s  =  |P_{1\bar{4}}P_{2\bar{3}}\rangle_s
 \ \ \ {\rm and} \ \ \
|2T\rangle_s  =  |{\bf V}_{1\bar{4}}
\cdot{\bf V}_{2\bar{3}}\rangle_s.
\end{equation}
And in the third channel:
\begin{equation}
\label{sp3}
|3S\rangle_s=|S_{12}S_{\bar{3}\bar{4}}\rangle_s
 \ \ \ {\rm and} \ \ \ |3T\rangle_s=
|{\bf A}_{12}\cdot{\bf A}_{\bar{3}\bar{4}}\rangle_s.
\end{equation}
In this notation $S_{ij}$ and ${\bf A}_{ij}$
stand for the scalar and axial
vector spin wave functions respectively, and the pseudoscalar and
vector spin wavefunctions $P_{ij}$ and ${\bf V}_{ij}$ are
defined in terms of
their linear combinations.

Except in the diagonal terms corresponding to the
$PP$ (i.e. pseudoscalar-pseudoscalar) sector
of the second gluonic channel to be
discussed below, the flavour wave functions are  taken to  be
trivial everywhere, just
giving rise to an isospin conserving
factor as the overlap of any two of them. ;
This happens in the absence of any mechanism for  flavour
change of a  quark or antiquark. Actually,  flavour changing is
possible in any channel through annihilation of quarks and
antiquarks of the same flavour. Our  consideration of
these processes in the
$PP$ sector of the second channel only is, therefore, an
approximation. This sector
is singled out because it is here that the annihilation effects are
apparently most significant and cannot be neglected  in any
realistic model of the processes involving mesons. This
is because they are
supposed to be responsible for the mass difference between
the (pseudoscalar) isoscalar and isovector mesons.
The annihilation effects are
negligible in the (spin) vector-vector sector of the second channel,
because of the small difference in mass of the spin one  isoscalar
and isovector  mesons i.e. $\omega$ and $\rho$.

When these annihilation processes are incorporated, the
flavour wave function gets
mixed with the quark position dependent part. This follows because
the size of
a mesonic cluster depends upon the masses
of the quarks (antiquarks) it
contains. Thus the $PP$ sector of the second channel is
considered separately below. For all other channels, we
can consider the quark
position dependent part of the wave function--referred to as
`quark wave function' in the following--separately from the flavour
part. This quark wave function is a function of four 3-vectors
${\bf r}_1,{\bf r}_2,{\bf r}_{\bar{3}} \ {\rm and}
\ {\bf r}_{\bar{4}}$.
These can be replaced by their combinations, with
one of them as the
overall centre-of-mass co-ordinate of the whole system
${\bf R}_{c}$ and
three others which
are taken, here, to be different in different channels. Writing
explicitly,  these are:

\noindent In the first channel
(with the gluonic part of the wave function
as $|1\rangle _g$)
\begin{equation}
\label{nvec1}
{\bf R}_1=\frac{{\bf r}_1+s{\bf r}_{\bar{3}}
-s{\bf r}_2-{\bf r}_{\bar{4}}}{1+s},
\ \ {\bf y}_1={\bf r}_1-{\bf r}_{\bar{3}} \ \ {\rm and} \ \ {\bf
z}_1={\bf r}_2-{\bf r}_{\bar{4}}.
\end{equation}
\noindent In the second channel
 \begin{equation}
\label{nvec2}
{\bf R}_2=\frac{{\bf r}_1+{\bf r}_{\bar{4}}-{\bf r}_2
-{\bf r}_{\bar{3}}}{2},
 \ \ {\bf y}_2={\bf r}_1-{\bf r}_{\bar{4}} \ \ {\rm and} \ \ {\bf
z}_2={\bf r}_2-{\bf r}_{\bar{3}}.
\end{equation}
\noindent And in the third channel
 \begin{equation}
\label{nvec3}
{\bf R}_3=\frac{{\bf r}_1+{\bf r}_2-{\bf r}_{\bar{3}}
-{\bf r}_{\bar{4}}}{1+s},
 \ \ {\bf y}_3={\bf r}_1-{\bf r}_2  \ \ {\rm and}  \ \ {\bf
z}_3={\bf r}_{\bar{3}}-{\bf r}_{\bar{4}}.
\end{equation}

The quark wave function in any channel is written, in the following,
as a product of two factors, one being a function of ${\bf R}_k$ and
the other of ${\bf y}_k$ and ${\bf z}_k$ only, for $k=1,2$ or $3$.
The former is denoted by
$\chi_{kI}({\bf R}_k)$, with $I$ designating the spin state
(singlet-singlet
or triplet-triplet),  and the latter by
$\xi_k({\bf y}_k)\zeta_k({\bf z}_k)$, with
$\xi_k({\bf y}_k)$ and $\zeta_k({\bf z}_k)$ corresponding to
the two mesonic clusters of the channel $k$. The
spatial dependence of these on ${\bf y}_k$ and ${\bf z}_k$ is taken
to be gaussian in consistency with the choice of
the quadratic form of
the inter-quark potential in eq.(\ref{nqd}).
But  the $\chi_{kI}\/$'s are treated as  variational
functions to be determined by solving the approximate
coupled Schr\"{o}dinger
equations, using the `resonating group method'   \cite{wildt77}.

 With the above mentioned forms  of the quark wave functions in
different channels, the
total state vector of the whole $q^2\bar{q}^2$ system is written as
\begin{equation}
\label{psi}
|\Psi(q_1,q_2,\bar{q}_3,\bar{q}_4;g)\rangle =
\sum_{kI} |k\rangle _g |kI\rangle_s|k\rangle_f\psi_{c}({\bf
R}_{c}) \chi_{kI}({\bf R}_k)\xi_k({\bf
y}_k)\zeta_k({\bf z}_k),
\end{equation}
with
\begin{equation}
\label{dsizek}
\xi_k({\bf y}_k)=\frac{1}{{(2\pi d_{k1}^2)}^{3/4}}{\rm exp}[-{\bf
y}_k^2/4d_{k1}^2] \ \ \ {\rm and}
\ \ \ \zeta_k({\bf z}_k)
=\frac{1}{{(2\pi d_{k2}^2)}^{3/4}}{\rm exp}[-{\bf
z}_k^2/4d_{k2}^2].
\end{equation}

This is actually the case with annihilation neglected.
In that situation
 the mesons represented by
$\xi_1, \zeta_1,\xi_3$ and $ \zeta_3$ have one light and one strange
(antistrange) particle. On the other hand, that denoted by $\xi_2$
has both
particles as the light ones (up and down) and $\zeta_2$
has both the quark and antiquark as the heavier ones.
It follows thus from the properties of the solutions of a 3-d
harmonic oscillator that $d_{11}, d_{12},d_{31}$ and
$d_{32}$ have a particular  value, say, $d'$,  $d_{21}$ has
a different one, say, $d$, and $d_{22}$
differs from all these having
a value denoted by $ d''$. Quantitatively
\begin{equation}
\label{drat}
\frac{{d'}^2}{d^2}=\sqrt{\frac{m(m_s+m)}{2mm_s}}
=\sqrt{\frac{s+1}{2s}}
\ \ \ {\rm and} \ \  \
\frac{{d''}^2}{d^2}=\sqrt{\frac{2m}{2m_s}}=\sqrt{\frac{1}{s}}.
\end{equation}
It must be pointed out that we are using
the approximation of neglecting the spin dependence of
the size of any
cluster. For the absolute magnitudes of the sizes, the relation
$d^2=\sqrt{3}R_n^2/2$ is used relating  the radius  of a meson
composed of the light mesons only to the r.m.s. charge radius
$R_n$=0.6 fm of a nucleon whose $qqq$ wave function is generated by
the same quadratic confining potential.

The $PP$  (or $S$) sector   of
the second channel needs special consideration
because of the annihilation and
creation processes, making the pairs
$(2,\bar{3})$ and $(1,\bar{4})$ mixtures
of  $s\bar{s}$,$u\bar{u}$ and $d\bar{d}$ in the flavour space. The
particular  combinations depend upon the physical mesons taking part
in the scattering process which may be $\eta$ and/or $\eta'$.
In the isovector sector only
$(2,\bar{3})$ has an amplitude for going
to, say, $u\bar{u}$ from its original
flavour state as $s\bar{s}$, resulting
in $\eta$ or $\eta'$. As the size, and hence
the quark wavefunction, of a particular
cluster is related to the masses of the
quark it contains, these amplitudes for
having different flavour contents imply
that in this particular channel we
cannot write the quark and flavour
wavefunctions separately as it is possible for the other
channels. Rather these two are mixed here, giving
us the combined quark-flavour
wavefunctions (except $\chi_{2S}({\bf R}_2)$)  as

\begin{equation}
\label{2sfq}
|2S\rangle_{fq}=|M_{1\bar{4}}\rangle_{fq}
|M_{2\bar{3}}\rangle_{fq}=\left\{
\begin{array}{ll}
|\eta'_{1\bar{4}}\rangle_{fq}|\eta'_{2\bar{3}}\rangle_{fq}
& \ \ \ \ \ {\rm
for} \ \ \ \ \
\eta'\eta' \ \ \ {\rm mesons} \\
|\eta_{1\bar{4}}\rangle_{fq}|\eta'_{2\bar{3}}\rangle_{fq}
& \ \ \ \ \ {\rm or} \\
|\eta'_{1\bar{4}}\rangle_{fq}|\eta_{ 2\bar{3}}\rangle_{fq}
& \ \ \ \ \ {\rm
for}\ \ \ \ \
\eta\eta'\  \ \ {\rm mesons} \\|\eta_{ 1\bar{4}}\rangle_{fq}|\eta_{
2\bar{3}}\rangle_{fq} & \ \ \ \ \ {\rm for}\ \ \ \  \
\eta\eta\ \ \  {\rm mesons} \\ |\pi_{ 1\bar{4}}\rangle_{fq}|\eta'_{
2\bar{3}}\rangle_{fq} & \ \ \ \ \ {\rm for}\ \ \ \ \  \pi\eta' \ \
 \ {\rm mesons} \\ |\pi_{ 1\bar{4}}\rangle_{fq}|\eta_{
2\bar{3}}\rangle_{fq} & \ \ \ \ \ {\rm for}\ \ \ \ \  \pi\eta
 \ \ \ {\rm mesons} \end{array}\right.,
\end{equation}
with

\begin{eqnarray}
|\eta_{ij}\rangle_{fq} & = &
\cos\theta\frac{|d\bar{d}+u\bar{u}\rangle_f}{\sqrt{2}}\xi_2({\bf
r}_{ij})-\sin\theta|s\bar{s}\rangle_f
\zeta_2({\bf r}_{ij}) \nonumber \\
\label{etfq}
|\eta'_{ij}\rangle_{fq} & = &
\sin\theta\frac{|d\bar{d}+u\bar{u}\rangle_f}{\sqrt{2}}\xi_2({\bf
r}_{ij})+\cos\theta|s\bar{s}\rangle_f\zeta_2({\bf r}_{ij}) \\
\label{pifq}
|\pi_{1\bar{4}}\rangle_{fq} & = &
|\pi\rangle_f\xi_2({\bf r}_{1\bar{4}})
=|\pi\rangle_f\xi_2({\bf y}_2).
\end{eqnarray}

\noindent Here
\begin{equation}
\label{xzr}
\xi_2({\bf r}_{ij})=\frac{1}{{(2\pi
d_{21}^2)}^{3/4}}\exp[-{\bf r}_{ij}^2/4d_{21}^2]
 \ \ \
{\rm and} \ \ \ \zeta_2({\bf r}_{ij})
=\frac{1}{{(2\pi d_{22}^2)}^{3/4}}
\exp[-{\bf r}_{ij}^2/4d_{22}^2],
\end{equation}
where $d_{21}=d,d_{22}=d''$,
and $\theta(=34.7^{\circ})$ is related to the mixing angle
$\theta_P=-20^{\circ}$ of flavour singlet
and octet resulting in $\eta$ and
$\eta'$ (see pages III.68-69 of   \cite{data92}).

After this discussion of the total wavefunction, we write the
Schr\"{o}dinger equation for the system in hand:
\begin{equation}
\label{nsch}
(H-E_c)|\Psi(q_1,q_2,\bar{q}_3,\bar{q}_4;g)\rangle =0
\end{equation}
where $H$ is the total hamiltonian  and $E_c$ is the total
centre-of-mass energy of the $q^2\bar{q}^2$ system.
This also means that the overlap of
$(H-E_c)|\Psi\rangle $  with an arbitrary
variation $|\delta\Psi\rangle $ of the state vector
$|\Psi\rangle $ vanishes.
In $|\delta\Psi\rangle $ we consider,
as in resonating group method calculations, only the
variations  in $\chi_{kI}$. Thus we write
\[\langle \delta\Psi|H-E_c|\Psi\rangle
=\sum_{kIlJ}\int d^3{\bf R}_cd^3{\bf
R}_kd^3{\bf y}_kd^3{\bf z}_k\]
\begin{equation}
\label{nvar}
\!\!\!\!\!\!\!\!\!\!\!\!\psi_{c}({\bf R}_{c})
\delta\chi_{kI}({\bf R}_k) \xi_k({\bf y}_k)\zeta_k({\bf
z}_k)\ _f\langle k| _s\langle kI|
_g\langle k|H-E_c|l\rangle _g |lJ\rangle_s
|l\rangle_f \psi_{c}({\bf R}_{c})\chi_{lJ}({\bf R}_l)
\xi_l({\bf y}_l)\zeta_l({\bf z}_l)=0.
\end{equation}
To do the four space integrations implied in
the overlap of the diquark
diantiquark position dependent
$(H-E_c)|\Psi\rangle $ and $|\delta\Psi\rangle $,
any of the three sets of
three-vectors defined by eqs.(\ref{nvec1}),(\ref{nvec2}) and
(\ref{nvec3}), along with ${\bf R}_c$, can be used. The choice ${\bf
R}_k,{\bf y}_k$ and ${\bf z}_k$ has, however, a clear advantage. The
arbitrary variations $\delta\chi_{kI}({\bf R}_k)$'s
for different (but
continuous!) values of
${\bf R}_k$, and of $k$ and $I$ as well,
are linearly independent and hence
their coefficients in
eq.(\ref{nvar}) should be  zero. With the trivial ${\bf
R}_c$ integration performed
to give a finite result using, say, box
normalization, this leads to

\begin{equation}
\label{czerg}
\!\!\!\!\!\!\!\!\!\!\!\!\sum_{lJ}\int d^3{\bf y}_kd^3{\bf z}_k
\xi_k({\bf y}_k)\zeta_k({\bf z}_k)\
_f\langle k|\,_s\langle kI|\,_g\langle
k|H-E_c|l\rangle_g|lJ\rangle_s|l\rangle_f\chi_{lJ}({\bf
R}_l)\xi_l({\bf y}_l)\zeta_l({\bf z}_l)=0,
\end{equation}
for $k,l$=1,2 or 3 and $I,J=S$ or $T$
(except the special case of $k,l$=2 and
$I,J=S$ where a mixed quark-flavour wavefunction is used).

In \cite{masu91} (and also in the previous section)
the suggested hamiltonian of the system  was mentioned in
the spin independent context. Despite the additional
degrees of freedom incorporated here we proceed in the same fashion,
considering first the hamiltonian
in the two-body potential model limit and
then modifying the off-diagonal elements. Now
the  spin dependent part of the hamiltonian,
composed of the terms
corresponding to hyperfine (contact as well as
tensor) and  spin-orbit
interactions, has to be considered as well, along with
that representing the annihilation effects. But our
job here is simpler  than this because of our constraint to the
S-wave ground states, implying that the only
additional spin-dependent term in the hamiltonian
to be dealt with is that
representing the hyperfine contact interaction.

For  the hyperfine term  in the two-body potential model limit
 we take  the expression given by one gluon exchange
(used, with some modifications, in   \cite{wn90}), and
sum over all the pairs:
\begin{equation}
\label{hyph}
V^{hyp}=\sum_{i<j}V^{hyp}_{ij}=-\sum_{i<j}{\bf F}_i\cdot{\bf
F}_j\frac{8\pi\alpha^{ij}_s}{3m_im_j}
\delta^3({\bf r}_{ij})\ {\bf S}_i\cdot{\bf S}_j .
\end{equation}
Numerical values of $\alpha^{ij}_s$, fitted to the light
meson spectroscopy
below, will be taken as varying with the sum of the masses
 of the particles $i$ and $j$; thus
each of them will eventually be
replaced by $\alpha_s^{ll},\alpha_s^{ls}$ or $\alpha_s^{ss}$,
$l$ standing for
a quark(antiquark) of light mass (i.e. up  or down) and $s$ for
a strange (heavier) one.

The pair annihilation and creation effects (where considered)
are represented by a hamiltonian term $V^a$
(denoted by $H_A$ in the
hamiltonian appearing in   \cite{godi85})
operating in the flavour space only. This should be a
sum of two terms
belonging to the pairs $ 1\bar{4}$ and $ 2\bar{3}$. Thus
\begin{equation}
\label{va12}
V^a=V^a_{1\bar{4}}+V^a_{2\bar{3}}.
\end{equation}
The matrix elements of $V_{ij}^a$ are written as
\begin{eqnarray}
_f\langle u\bar{u}|V^a_{ij}|u\bar{u}\rangle_f & = & \ _f\langle
d\bar{d}|V^a_{ij}|u\bar{u}\rangle_f=\ _f\langle
d\bar{d}|V^a_{ij}|d\bar{d}\rangle_f= l, \\
\label{vaijd}
_f\langle u\bar{u}|V^a_{ij}|s\bar{s}\rangle_f & = & \ _f\langle
d\bar{d}|V^a_{ij}|s\bar{s}\rangle_f
=\sqrt{l}\sqrt{n} \ \ \ \ \ {\rm and} \
\ \ \ \ _f\langle s\bar{s}|V^a_{ij}|s\bar{s}\rangle_f=n. \ \ \ \ \
\ \ \ \ \  \
\end{eqnarray}
in the corresponding flavour spaces.
The mass (or flavour)  dependence shown here
is in qualitative agreement  with the mass dependence of the
annihilation term  $H_A$ appearing in   \cite{godi85}.
In the above, $l$ and $n$ are phenomenological parameters
to be fitted to the
masses and flavour wave functions of  $\pi$, $\eta$
and $\eta'$ mesons.

This specification of the hamiltonian of the system means that
$H$ in eq.(\ref{czerg}) is to be replaced
by $K+V^p+V^a+\sum_{i=1}^{\bar{4}}m_i$, with
$V^p=V^{cf}+V^{hyp}$ i.e. a sum of the confinement and
hyperfine potentials.
Using eqs.(\ref{nvform}) and (\ref{hyph}) for $V^{cf}$ and $V^{hyp}$
respectively, it can be seen easily that now the matrix elements of
$V^p$ in our spin basis, in the two-body potential model,
would be given by
\begin{eqnarray}
\label{vpform}
_s\langle X|V^p|X'\rangle_s & = & \sum_{i<j}{\bf F}_i\cdot{\bf F}_j\
(V_{ij})_{X_s,X'_s},    \ \ \ \ \  \ \ \ \ \  \ \ \ \ \
 \ \ \ \ \ {\rm with}
\\
\label{vijxx}
(V_{ij})_{X_s,X'_s} & = & v_{ij}\ _s\langle X|X'\rangle_s
-\frac{8\pi \alpha^{ij}_s}{3m_im_j}\delta^3({\bf
r}_{ij})\ _s\langle X|{\bf S}_i\cdot{\bf S}_j|X'\rangle_s.
\end{eqnarray}

 In this form, $_s\langle X|V^p|X'\rangle_s$ is very  similar to
the expression for $V$ appearing in eq.(\ref{nvform}). So, as far as
its matrix elements between the gluonic states are concerned, the
whole formalism developed in \cite{masu91} (also reported in
the previous section) can be utilized, provided
the spin state dependence of the matrix coefficients $V_{ij}$,
replacing
$v_{ij}$, of  ${\bf F}_i\cdot{\bf F}_j$  is taken care of.
Thus, the matrix element of the $V^p$ term in the
hamiltonian  between any two gluonic-spin
states appearing in eq.(\ref{czerg})
 is given by (see eq.(\ref{nvfom}) for comparison)
\begin{equation}
\label{vpgs}
V^p=  \left( \begin{array}{ccl}
-\frac{4}{3}(V_{1\bar{3}}+V_{2\bar{4}})_{1,1} &
\frac{4f}{9}\left( \begin{array}{c} V_{12}+V_{\bar{3}\bar{4}}
\\ -V_{1\bar{3}}-V_{2\bar{4}} \\
-V_{1\bar{4}}-V_{2\bar{3}} \end{array} \right)_{1,2} &
\frac{2f}{3\sqrt{3}}\left( \begin{array}{c}
-2(V_{1\bar{3}}+V_{2\bar{4}}) \\ +V_{1\bar{4}}
+V_{2\bar{3}} \\
-V_{12}-V_{\bar{3}\bar{4}} \end{array} \right)_{1,3} \\  &
-\frac{4}{3}(V_{1\bar{4}}
+V_{2\bar{3}})_{2,2} & \frac{2f}{3\sqrt{3}}\left(
\begin{array}{c} 2(V_{1\bar{4}}+V_{2\bar{3}}) \\ +V_{12}
+V_{\bar{3}\bar{4}} \\
-V_{2\bar{4}}-V_{1\bar{3}} \end{array} \right)_{2,3} \\
{\rm symmetric} &  &
\begin{array}{l} -\frac{1}{3}\left( \begin{array}{c}
2(V_{12}+V_{\bar{3}\bar{4}}) \\ +V_{1\bar{3}}
+V_{2\bar{4}} \\ +V_{1\bar{4}}+V_{2\bar{3}} \end{array}
\right)_{3,3} \\ +\frac{5}{2}I\,Df(1-f) \end{array}
\end{array} \right),
\end{equation}
in the gluonic-spin basis $|1\rangle_g|1S\rangle_s$,
$|1\rangle_g|1T\rangle_s,
\cdots,|3\rangle_g|3T\rangle_s$.

The above matrix should
give $6\times 6=36$ matrix elements of the $V^p$ operator in  its
6-dimensional basis. But as written above, it has only {\em nine}
elements.
Actually, every term in every  matrix element of the above is meant
to stand for a $2\times 2$ matrix defined by
\begin{equation}
\label{one22}
(V_{ij})_{k,l}=\left(\begin{array}{cc} (V_{ij})_{kS,lS} &
(V_{ij})_{kS,lT} \\
(V_{ij})_{kT,lS} & (V_{ij})_{kT,lT}\end{array}\right),
\end{equation}
for $k,l$=1,2 or 3. $(V_{ij})_{kI,lJ}$ is given through
eq.(\ref{vijxx}) and
$I$ multiplying the $D$ term in 3,3 element of
eq.(\ref{vpgs}) is a $2\times 2$ identity matrix.
To determine the overlap of any two spin states and
the corresponding matrix
element of the operator ${\bf S}_i\cdot{\bf S}_j$, use has
to be made of
the definitions
given through eqs.(\ref{sp1}),(\ref{sp2}) and (\ref{sp3}) above.
The results
are reported in Appendix A.

As far as the spin dependence is concerned,
the other terms in the hamiltonian are unit operators.
Using the results reported in the previous section
for the matrix elements
between gluonic states, we write eq.(\ref{czerg}) as
\begin{eqnarray}
\sum_{lJ}\int d^3{\bf R}_l
\left[{\cal K}_{kI,lJ}({\bf R}_k,{\bf R}_l)+{\cal
V}^{cf}_{kI,lJ}({\bf R}_k,{\bf R}_l)+{\cal
V}^{hyp}_{kI,lJ}({\bf R}_k,{\bf R}_l)
\phantom{\sum_i^{\bar{4}}}\right.   & & \nonumber \\
\label{master}
  \left. -\left(E_c-\sum_{i=1}^{\bar{4}}m_i\right){\cal
N}_{kI,lJ}({\bf R}_k,{\bf R}_l)\right]\chi_{lJ}({\bf R}_l)
& = & 0, \ \ \ \ \ \ \ \
\end{eqnarray}
for $l=1,2 $ and 3 along with $J=S$ and $T$. This gives six equations,
each for one of the six possible values of the pair of
variables $k,I$ with $k=1,2 $ or 3  and $I=S$ or $T$.
Here ${\cal K}_{kI,lJ},{\cal
V}^{cf}_{kI,lJ},{\cal V}^{hyp}_{kI,lJ}$ and ${\cal N}_{kI,lJ}$
are defined
through the following equations:
\begin{equation}
\label{gcalkd}
\int d^3{\bf R}'_l{\cal K}_{kI,lJ}
({\bf R}_k,{\bf R}'_l)\chi_{lJ}({\bf R}'_l)
=\int d^3{\bf y}_kd^3{\bf z}_k
\xi_k({\bf y}_k)\zeta_k({\bf z}_k)K_{kI,lJ}\chi_{lJ}({\bf
R}_l)\xi_l({\bf y}_l)\zeta_l({\bf z}_l) \ \ \ \ \
\end{equation}
\begin{equation}
\label{gcalvcd}
\int d^3{\bf R}'_l{\cal
V}^{cf}_{kI,lJ}({\bf R}_k,{\bf R}'_l)
\chi_{lJ}({\bf R}'_l)=\int d^3{\bf y}_kd^3{\bf z}_k
\xi_k({\bf y}_k)\zeta_k({\bf z}_k)V^{cf}_{kI,lJ}\chi_{lJ}({\bf
R}_l)\xi_l({\bf y}_l)\zeta_l({\bf z}_l) \ \ \ \ \
\end{equation}
\begin{equation}
\label{gcalvhd}
\int d^3{\bf R}'_l{\cal
V}^{hyp}_{kI,lJ}({\bf R}_k,{\bf R}'_l)
\chi_{lJ}({\bf R}'_l)
=\int d^3{\bf y}_kd^3{\bf z}_k\xi_k({\bf y}_k)
\zeta_k({\bf z}_k)V^{hyp}_{kI,lJ}\chi_{lJ}({\bf
R}_l)\xi_l({\bf y}_l)\zeta_l({\bf z}_l) \ \ \ \ \
\end{equation}
\begin{equation}
\label{gcalnd}
\int d^3{\bf R}'_l{\cal
N}_{kI,lJ}({\bf R}_k,{\bf R}'_l)\chi_{lJ}({\bf R}'_l)=
\int d^3{\bf y}_kd^3{\bf z}_k\xi_k({\bf y}_k)
\zeta_k({\bf z}_k)N_{kI,lJ}\chi_{lJ}({\bf
R}_l)\xi_l({\bf y}_l)\zeta_l({\bf z}_l), \ \ \ \ \
\end{equation}
with $K_{kI,lJ},V^{cf}_{kI,lJ}$ and $V^{hyp}_{kI,lJ}$  representing
the matrix elements of the $K,V^{cf}$ and $V^{hyp}$ operators between
the spin and  gluonic states
appearing in eq.(\ref{czerg}). $N_{kI,lJ}$ is the
overlap of these states, calculated using  the results mentioned in
the previous section
along with those in Appendix A for the spin overlap factor $_s\langle
kI|lJ\rangle_s$.

The spatial integrations appearing on the RHS of
the eqs.(\ref{gcalkd}) to
(\ref{gcalnd}) are done after substituting the
expressions for $K_{kI,lJ},V^{cf}_{kI,lJ},
V^{hyp}_{kI,lJ}$ and $N_{kI,lJ}$,
obtained through the procedure outlined above.
In these calculations
the diagonal ($k=l$) and off-diagonal ($k\neq l$)
cases were dealt with
separately. In the former
case, $\chi_{lJ}({\bf R}_l)$ is linearly independent
of the integration
variables ${\bf y}_k$
and ${\bf z}_k$ and thus was simply taken out of the integrations.
In the case of off-diagonal terms (with $k\neq l$),
the integration variables
(${\bf y}_k$ and ${\bf z}_k$) were replaced by their
equivalent
combinations with one identical to ${\bf R}_l$,
and the other one independent
of it. Integrating out the vector  independent of ${\bf R}_l$,
expressing the remaining one  on the RHS
(apart from    ${\bf R}_l$) in terms of
${\bf R}_k$ and ${\bf R}_l$, we got the results for
${\cal K}_{kI,lJ},{\cal
V}^{cf}_{kI,lJ},{\cal V}^{hyp}_{kI,lJ}$ and ${\cal N}_{kI,lJ}$
after comparing with the LHS of the corresponding equation.

Where annihilation is considered,
we have to use the combined quark and flavour wave function
$|2S\rangle_{fq}$ defined through eq.(\ref{2sfq}).
Thus in the  diagonal term corresponding to the
$2S$  channel we have, in place of the $k=l=2$ and $I=J=S$ term in
eq.(\ref{czerg}),
\[\int d^3{\bf r}_{1\bar{4}}d^3{\bf r}_{2\bar{3}}\ _{fq}\langle 2S|\,
_s\langle
2S| _g\langle 2|H-E_c|2\rangle_g|2S\rangle_s|2S\rangle_{fq}\
\chi_{2S}({\bf R}_2)\]
\begin{eqnarray}
& = & \int d^3{\bf r}_{1\bar{4}}d^3{\bf r}_{2\bar{3}}
\ _{fq}\langle 2S|
\,_s\langle
2S| _g\langle 2|K+V^p-(E_c-\sum_{i=1}^{\bar{4}}m_i)
|2\rangle_g|2S\rangle_s
|2S\rangle_{fq}\chi_{2S}({\bf
R}_2) \nonumber \\
\label{2dg}
& + & \int d^3{\bf r}_{1\bar{4}}d^3{\bf r}_{2\bar{3}}\
_{fq}\langle 2S|V^a|2S\rangle_{fq}\chi_{2S}({\bf R}_2).
\end{eqnarray}
As expressed through eq.(\ref{2sfq}), the form  of $|2S\rangle_{fq}$
depends upon the physical content of the $2S$  channel.
This would result in different expressions for each of
${\cal K}_{2S,2S},{\cal V}^{cf}_{2S,2S},{\cal V}^{hyp}_{2S,2S}$ and
${\cal N}_{2S,2S}$ for different pairs of mesons in the channel.
In the following calculations we restrict ourselves to
just the lowest,
in threshold energy, channels:
$\eta\eta$ in the isoscalar, plus  $\eta\pi$ and $\eta'\pi$ in the
the isovector sector. This is done because of our
special interest in the
behaviour of the $K\bar{K}$ \label{`etapneglect'}
system near the threshold (see the next section).

The results thus obtained for these
non-local kernels, for all the values of $k,l,I$ and $J$,
appearing in
eq.(\ref{master}) are reported in Appendix B. Substitution of
these in  eq.(\ref{master}) would
gives {\em six} coupled equations. However we neglect
all the connections to the third gluonic channel, justified to
some extent  by the absence of any significant effect
of removing  the
third channel in the spinless case (see fig.4 of   \cite{masu91}).
This leaves us with just {\em four} equations (two of these
are written below
as eq.(\ref{cempkkc}) and (\ref{cempeec})). The off diagonal
(i.e. with
$k\neq l$) terms in these equations
tend to zero for large inter-cluster
distances. Thus for consistency with the observed
meson spectroscopy
we would require the constant term in each of
the diagonal parts to be identical to
the sum of  masses of the mesons
present in the corresponding channel.
Fitting, in this way, to the masses of
$K,\eta,\eta',\pi,K^{\ast},\omega$
(or $\rho$)
and $\phi$ mesons, we get the following values of
the above mentioned
free parameters of the formalism:
\begin{eqnarray}
m & = & 277 \ {\rm MeV}, \ \ \ \ \
s  =  1.955, \ \ \ \ \  \bar{C}  =  456 \ {\rm MeV}, \nonumber \\
\alpha^{ll}_s & = & 1.583, \ \ \ \ \ \ \ \
\alpha^{ls}_s  =  1.561, \ \ \ \ \ \alpha^{ss}_s =
1.501, \nonumber \\
l & = & 272 \ {\rm MeV} \ \ \ \ \ \ {\rm and}
\ \ \ \ \ \ n  =  67.4 \ {\rm MeV}.
\end{eqnarray}
For $C$ (see eq.(\ref{nqd})),
the  equality of kinetic and potential energies of a harmonic
oscillator is used, giving us
\begin{equation}
C=-\frac{1}{4d^2}\frac{3}{4}\omega_2^l
=-\frac{3}{16md^4}=-270 \ {\rm
MeV/{fm}^2}.
\end{equation}

After this parameter fit (except for $\bar{k}$ to be discussed in the
next section), we write down {\em two} of the {\em four} coupled
equations mentioned above. The remaining two equations would involve
vector mesons. These are not incorporated
beyond this stage because of our
above mentioned neglect of channels opening at
energies significantly higher  than the $K\bar{K}$ threshold.
\begin{eqnarray}
 & & \left[M_K+M_{\bar{K}}
-\frac{1}{2\mu_{K\bar{K}}}\underline{\nabla}_{{\bf
R}_1}^2-E_c\right]\chi_{1S}({\bf R}_1)    \nonumber \\
& & +e_0\int d^3{\bf
R}_2\left\{\!\!\!\!\!\!\!\!\!
\phantom{\sum_1^{\bar{4}}}
\left[-\frac{1}{2m}\frac{1}{6}\left(q_{11}{\bf R}_1^2+q_{12}{\bf
R}_2^2+q_{10}\right)+C({\bf R}_1)\right]\exp\left[-e_2{\bf
R}^2_2-e_1{\bf R}_1^2\right]
\phantom{\frac{8}{3}}\right.  \nonumber \\
\label{cempkkc}
& &  \left.\!\!\!\!\!\!\phantom{\frac{8}{3}}
-G({\bf R}_1,{\bf R}_2)H       \!\!\!\!\!\!\!
\phantom{\sum_1^{\bar{4}}}
\right\}\chi_{2S}({\bf R}_2)  =  0, \ \ \ \ \ \ \ \ \
\ \ \ \ \ \ {\rm and} \\
 & & \left[M_a+M_b -\frac{1}{2\mu_{ab}}\underline{\nabla}_{{\bf
R}_2}^2 -E_c\right]\chi_{2S}({\bf R}_2)    \nonumber \\
 & & +e_0\int d^3{\bf
R}_1\left\{\!\!\!\!\!\!\!\!\!\phantom{\sum_1^{\bar{4}}}
\left[-\frac{1}{2m}\frac{1}{6}\left(q_{21}{\bf R}_1^2+q_{22}{\bf
R}_2^2+q_{20}\right)+C({\bf R}_1)\right]\exp\left[-e_2{\bf
R}^2_2-e_1{\bf R}_1^2\right]
\phantom{\frac{8}{3}}\right.  \nonumber \\
\label{cempeec}
& &  \left.\!\!\!\!\!\!\phantom{\frac{8}{3}}
-G({\bf R}_1,{\bf R}_2)H       \!\!\!\!\!\!\!
\phantom{\sum_1^{\bar{4}}}
\right\}\chi_{1S}({\bf R}_1)  =  0,
\end{eqnarray}
for $K\bar{K}\leftrightarrow ab$,
where $a$ and $b$ are the two
mesons in the second channel ($\eta\eta,\eta\pi$ or $\eta'\pi$).
$C({\bf R}_1)$ and $G({\bf R}_1,{\bf R}_2)$ appearing
in the above are
\pagebreak[3]
\begin{eqnarray}
C({\bf R}_1) & = & \frac{1}{2}\left(b_1{\bf R}_1^2+b_0\right)
-\frac{1}{6}\left(E_c+\frac{8}{3}\bar{C}
-2m(s+1)\right)    \\
G({\bf R}_1,{\bf R}_2) & = & l_{10}\exp\left[-(e_1+e'_1-l_{11}){\bf
R}^2_1-(e_2+l_{12}){\bf R}_2^2\right]\times  \nonumber \\
 & & \left[\alpha^{ls}_s\exp\left(l_{13}{\bf R}_1\cdot{\bf
R}_2\right)+\alpha^{ls}_s\exp\left(-l_{13}{\bf
R}_1\cdot{\bf
R}_2\right)\right]  \nonumber \\
 & & +l_{20}\exp\left[-(e_1+e_1'+l_{21}){\bf
R}_1^2-e_2{\bf R}_2^2\right]\times \nonumber \\
 & & \left[\alpha^{ll}_s\exp\left(l_{22}{\bf
R}_1^2\right)+s\alpha_s^{ss}\exp\left(-l_{22}{\bf
R}_1^2\right)\right]  \nonumber \\
 & &  +l_{30}\exp\left[-(e_1+e_1'+l_{31})
{\bf R}_1^2-(e_2+l_{32}){\bf
R}_2^2\right]\times \nonumber \\
\label{gr1r2def}
 & & \left[\alpha^{ls}_s\exp\left(l_{33}{\bf R}_1\cdot{\bf
R}_2\right)+\alpha^{ls}_s\exp\left(-l_{33}{\bf
R}_1\cdot{\bf R}_2\right)\right].
\end{eqnarray}
Moreover,
\begin{equation}
\label{caphd}
H=\frac{1}{6}\frac{8\pi}{3m^2s}
\frac{(2\kappa)^{3/2}}{(2\pi d^2)^{3/2}}.
\end{equation}
It should be noted that
in the coefficients of $\underline{\nabla}^2_{{\bf R}_1}$
and  $\underline{\nabla}^2_{{\bf R}_2}$,  the reduced masses of the
two pseudoscalar mesons of  the particular channel now appear.
This is done so as to ensure that the terms
involving $\underline{\nabla}^2_{{\bf R}_1}$
or  $\underline{\nabla}^2_{{\bf R}_2}$  give the correct kinetic
energy of the relative motion of the interacting physical mesons.
Other symbols appearing in the
above equations are defined in Appendix B
at appropriate places.

The kernels of the integrals appearing in the off-diagonal parts in
the above coupled equations
contain non-separable parts \newline $\exp(l_{13}{\bf R}_1\cdot{\bf
R}_2)\cdots$
$\exp(l_{33}{\bf R}_1\cdot{\bf R}_2)$. The presence of these
``non-separable potential'' terms makes the solution of the coupled
equations rather involved. To avoid that complication, we can solve
our problem  in the approximation of replacing these terms by their
truncated expansions which would leave us with
an inexact but easily manageable form of
the equations. With that strategy in
mind, the above equations were solved
first for the case of no hyperfine
interaction (thus avoiding non-separable terms)
by setting $H=0$ in the above
two coupled equations. The method used for
that is explained below for the
full interaction  case.
The resulting phase shifts for this no hyperfine case
(reported  in the next section)
are so small that it would be  a good
approximation  to take the variational wave functions in the
{\em absence}
 of hyperfine interaction
as the wave-functions corresponding to a freely propagating plane
wave. Using this approximation for
 the variational wave functions $\chi_{1S}({\bf
R}_1)$ and $\chi_{1S}({\bf R}_1)$  even in the {\em presence} of
hyperfine
interaction, we looked for a reasonable separable
approximation to our non-separable potential terms.
As far as the terms $\exp(l_{13}{\bf
R}_1\cdot{\bf R}_2)$ and $\exp(-l_{13}{\bf R}_1\cdot{\bf R}_2)$ are
concerned, it was seen to be a very good approximation to just
replace them with the exponential expansion up to the second power in
$l_{13}{\bf R}_1\cdot{\bf R}_2$. But the terms multiplying $l_{30}$
are not so easy to manage. Their separable approximation involved two
parameters which had to be adjusted for each of value of the
kinetic energy in the centre of mass frame. Written explicitly, our
approximation has been  to take
\pagebreak[3]
\[ l_{30}\exp\left[-(e_1+e_1'+l_{31}){\bf R}_1^2-(e_2+l_{32}){\bf
R}_2^2\right]\times\]
\[\left\{\alpha^{ls}_s\exp\left(-l_{33}{\bf R}_1\cdot{\bf
R}_2\right)+\alpha_s^{ls}\exp\left(l_{33}{\bf R}_1\cdot{\bf
R}_2\right)\right\}\approx \]
\[ n_1l_{30}\alpha_s^{ls}\left\{\exp\left[-\tau_1\left(e_1+e'_1
+l_{31}+
e_2+l_{32}-l_{33}\right)\left({\bf R}_1^2+{\bf R}_2^2
\right)\right]\right. \]
\begin{equation}
\label{approx}
\left.-\left[-\tau_1\left(e_1+e'_1+l_{31}+
e_2+l_{32}+l_{33}\right)\left({\bf R}_1^2+{\bf R}_2^2\right)
\right]\right\},
\end{equation}
in $G({\bf R}_1,{\bf R}_2)$, and hence in the above two coupled
equations.
Here $n_1$ and $\tau_1$ are the above mentioned two energy dependent
parameters. Actually these are to be
used in eq.(\ref{cempkkc}) {\em only}. Those appearing in
eq.(\ref{cempeec}) are different
and, therefore, are denoted by $n_2$ and $\tau_2$ instead.

Both sides of eq.(\ref{approx}), multiplied by $R_2^2\chi_{2S}(R_2)$
and
integrated over $R_2$, were plotted as functions of $R_1$. This
comparison showed that in this way
even the worst discrepancy could be reduced to less than
10 percent of the total hyperfine coupling for that particular value
of $R_1$ and the on-shell momentum $p_c(2)$. After calculating the
$T$  and $S$
scattering matrices, the results were  checked for any possible
deviation from  unitarity of the $S$ matrix and symmetry of
the $T$ matrix
(required by ``reciprocity'' of inelastic scattering, see, for
example, p.528 of   \cite{blatt}). The discrepancy was for
some cases as
bad as 30 percent, implying that the above approximation
needs to be improved by, for example, iterating it many times.
This improvement
remains to be made, although  it can be easily shown that
this re-adjusting of
the values of $n_1,n_2,\tau_1$ and $\tau_2$
(with improved functional
dependences of $\chi_{1S}({\bf R}_1)$ and $\chi_{2S}({\bf R}_2$))
is not needed for the range  of
energy where our immediate interest lies (e.g. below the $K\bar{K}$
threshold
in the first channel).  This follows because the formal momentum
space solutions (see eqs.(\ref{gsole11})
and (\ref{gsole12}) below) of the above
coupled equations, for that range of energy, would be of the form
\[\chi(p)=-\frac{1}{\Delta (p)}\times {\rm Constant}.\]
Changing values of $n_1,\tau_1,n_2$ and $\tau_2$  in that situation
would just affect the constant
coefficients of $\frac{1}{\Delta_1 (p_1)}$ and
$\frac{1}{\Delta_2 (p_2)}$, leaving
the momentum, and hence the space,
dependence of the solutions $\chi_{1S}$ and
$\chi_{2S}$ unchanged.

With the above approximation the integrand appearing in the coupling
terms in both of our coupled equations (\ref{cempkkc}) and
(\ref{cempeec}) are products of two factors,
each of them is a function of  $R_1$ {\em or}  $R_2$.
This means that in this form the two coupled equations can be solved
exactly, using the method demonstrated in appendix B of
\cite{masu91}. The procedure used in the present
case was to first write eqs.(\ref{cempkkc}) and
(\ref{cempeec}), in the approximate form
(see eq.(\ref{approx})), in momentum space.
For incoming waves in the first channel,
the formal momentum space solution  of these equations would be

\begin{eqnarray}
\chi_{1S}(p_1) & = & \frac{\delta(p_1-p_c(1))}{p_c^2(1)} \nonumber \\
& & -\frac{1}{\Delta_1(p_1)}\left[Q_1^{(1)}A_2(e_2)+Q_2^{(1)}B_2(e_2)
+Q_3^{(1)}A_2(e_2+l_{12})+Q_4^{(1)}B_2(e_2+l_{12})\right. \nonumber
\\ & &
+Q_5^{(1)}A_2(\tau_1\overline{e_1+e'_1+l_{31}+e_2+l_{32}-l_{33}})
\nonumber \\
& &
\left.+Q_6^{(1)}A_2(\tau_1\overline{e_1+e'_1+l_{31}+e_2+l_{32}
+l_{33}})
\right]
\label{gsole11} \\
\chi_{2S}(p_2) & = &
-\frac{1}{\Delta_2(p_2)}\left[Q_1^{(2)}A_1(e_1)+Q_2^{(2)}B_1(e_1)
\right. \nonumber \\
& & +Q_3^{(2)}A_1(e_1+e'_1+l_{21}-l_{22})+Q_4^{(2)}
A_1(e_1+e'_1+l_{21}+l_{22})
\nonumber \\
& & +Q_5^{(2)}A_1(e_1+e'_1-l_{11})+Q_6^{(2)}
B_1(e_1+e'_1-l_{11}) \nonumber \\
& & +Q_7^{(2)}A_1(\tau_2\overline{e_1+e'_1+l_{31}+e_2+l_{32}-l_{33}})
\nonumber \\
& &
\left.+Q_8^{(2)}A_1(\tau_2\overline{e_1+e'_1+l_{31}+e_2+l_{32}
+l_{33}})\right].
\label{gsole12}
\end{eqnarray}
The new symbols appearing in these equations are defined in Appendix
C.

\begin{sloppypar}
It is to be noted that in the
above equations ${\bf p}_1$ and ${\bf p}_2$
have been replaced
everywhere by $p_1$ and $p_2$ respectively, utilizing
the spherical
symmetry of our problem. Multiplying eq.(\ref{gsole11}) by
$ p_1^2 F_a\left(p_1,e_1\right),  p_1^2 F_b\left(p_1,e_1\right)$,
$ p_1^2 F_a\left(p_1,e_1+e'_1+l_{21}-l_{22}\right),
 p_1^2 F_a\left(p_1,e_1+e'_1+l_{21}+l_{22}\right),
 p_1^2 F_a\left(p_1,e_1+e'_1-l_{11}\right)$,
$ p_1^2 F_b\left(p_1,e_1+e'_1-l_{11}\right),
p_1^2 F_a\left(p_1,\tau_2\overline{e_1+e'_1+l_{31}+
e_2+l_{32}-l_{33}}\right)$ and
$ p_1^2 F_a\left(p_1,\tau_2\overline{e_1+e'_1+l_{31}+
e_2+l_{32}+l_{33}}\right)$ in turn and integrating w.r.t. $p_1$ gives
us 8 equations ($F_a(p_1,x)$ and $F_b(p_1,x)$ are the Fourier
transforms of
$\exp\left[-xR_1^2\right]$ and
$R_1^2\exp\left[-xR_1^2\right]$
respectively). Similarly multiplying eq.(\ref{gsole12}) by
the Fourier transforms
$ p_2^2 F_a\left(p_2,e_2\right),
 p_2^2 F_b\left(p_2,e_2\right),
 p_2^2 F_a\left(p_2,e_2+l_{12}\right)$,
$ p_2^2 F_b\left(p_2,e_2+l_{12}\right),
 p_2^2 F_a\left(p_2,\tau_1\overline{e_1+e'_1+l_{31}+
e_2+l_{32}-l_{33}}\right)$ and
$ p_2^2  F_a\left(p_2,\tau_1\overline{e_1+e'_1+l_{31}+
e_2+l_{32}+l_{33}}\right)$ and integrating w.r.t. $p_2$ gives
us 6 more equations.
These 14 equations can be written as a  matrix equation
\end{sloppypar}
\begin{equation}
\label{gqu1u2}
{\sf Q}{\sf U}_1={\sf U}_2
\end{equation}
with
\begin{equation}
\label{gu21def}
{\sf U}_2 =  4\pi\left(\begin{array}{c} F_a\left(p_c(1),e_1\right)
\\ F_b\left(p_c(1),e_1\right) \\
F_a\left(p_c(1),e_1+e'_1+l_{21}-l_{22}\right) \\
F_a\left(p_c(1),e_1+e'_1+l_{21}+l_{22}\right) \\
F_a\left(p_c(1),e_1+e'_1-l_{11}\right) \\
F_b\left(p_c(1),e_1+e'_1-l_{11}\right) \\
F_a\left(p_c(1),\tau_2\overline{e_1+e'_1+l_{31}+
e_2+l_{32}-l_{33}}\right) \\
F_a\left(p_c(1),\tau_2\overline{e_1+e'_1+l_{31}+
e_2+l_{32}+l_{33}}\right) \\
0 \\ 0 \\ 0 \\ 0 \\ 0 \\ 0 \end{array}\right),
\end{equation}
${\sf Q}$ a $14\times 14$ matrix containing many integrals,
and ${\sf U}_1$ a vector containing
$A_2(e_2),B_2(e_2),\cdots
,A_2(\tau_1\overline{e_1+e'_1+l_{31}+ e_2+l_{32}+l_{33}})$ \newline
and $A_1(e_1), B_1(e_1),\cdots,
A_1(\tau_2\overline{e_1+e'_1+l_{31}+ e_2+l_{32}+l_{33}})$
as its elements.
Inverting the matrix ${\sf Q}$ gives these 14 elements of the
${\sf U}_1$ vector.
With these values in hand, all the quantities in the expressions
for the variational functions $\chi_{1S}(p_1)$ and $\chi_{2S}(p_2)$
are known. So these can be now simply
obtained by making the usual replacement
of $p_1$ and $p_2$   by their on-shell values $p_c(1)$ and $p_c(2)$,
defined by eqs.(\ref{pc1def}) and (\ref{pc2def}), respectively
in eqs.(\ref{gsole11}) and (\ref{gsole12}).

{}From eqs.(\ref{gsole11}) and (\ref{gsole12}) the two $T$
matrix elements
$T_{1,1}$ and $T_{2,1}$, proportional to
the coefficients of the non-relativistic Green operators
$-\frac{1}{\Delta_1(p_1)}$ and $-\frac{1}{\Delta_2(p_2)}$
respectively,
can be read off. These are reported in Appendix D.
Similarly,  for incoming waves in channel 2, the use of ${\sf U}_2$
as
\begin{equation}
\label{gu22def}
{\sf U}_2  =  4\pi\left(\begin{array}{c} 0 \\ 0 \\
0 \\ 0 \\ 0 \\ 0 \\ 0 \\ 0 \\ F_a\left(p_c(2),e_2\right)
\\ F_b\left(p_c(2),e_2\right) \\
F_a\left(p_c(2),e_2+l_{12}\right) \\
F_b\left(p_c(2),e_2+l_{12}\right) \\
F_a\left(p_c(2),\tau_1\overline{e_1+e'_1+l_{31}+
e_2+l_{32}-l_{33}}\right) \\
F_a\left(p_c(2),\tau_1\overline{e_1+e'_1+l_{31}+
e_2+l_{32}+l_{33}}\right)  \end{array}\right),
\end{equation}
instead of that given by eq.(\ref{gu21def}), gives the two $T$ matrix
elements $T_{2,2}$ and $T_{1,2}$ reported also in Appendix D.

For the total energy in the centre of mass frame above the higher
threshold, both of the channels would be
open. Thus for incoming waves in either of them, there would be a
loss of flux in the incoming channel (with the total flux remaining
conserved).
Representing this {\em inelasticity} by a factor $\epsilon_k$, for
$k=1$ or
2, we can write
\begin{eqnarray}
\label{eps1d1}
S_{1,1} & = & 1-2iT_{1,1}=\epsilon_1e^{2i\delta_1} \\
\label{eps2d2}
S_{2,2} & = & 1-2iT_{2,2}=\epsilon_2e^{2i\delta_2}.
\end{eqnarray}
For elastic scattering,
$\epsilon_1$ or $\epsilon_2$ would be unity
for  incoming waves in  channel 1 or 2 respectively.

Below the lower threshold the situation would be qualitatively
different as,
with both $p_c(1)$ and
$p_c(2)$ being imaginary, $\delta(p_1-p_c(1))$  and
$\delta(p_2-p_c(2))$
would not contribute to the integration over all the
real values of
$p_1$ and $p_2$ performed to arrive at eq.(\ref{gqu1u2}).
Thus all the terms
collected in the vector ${\sf U}_2$ would be absent,
leaving us instead with
\begin{equation}
\label{qu1zero}
{\sf Q}{\sf U}_1=0.
\end{equation}
A non-trivial solution of this
equation for the elements of the vector
${\sf U}_1$ requires
\begin{equation}
\label{detqzero}
{\rm det}\:{\sf Q}=0.
\end{equation}
giving us a condition for the existence of  a bound state of the
whole system.

\section{Results}
\setcounter{equation}{0}
As should be clear from the expressions reported in Appendix B,
our coupled equations can describe a number of meson-meson
systems. Amongst these systems we have chosen
$K\bar{K}$, keeping in mind that it has been investigated by other
groups
using different models for the quark-quark  interaction.
An important issue is
whether the whole $K\bar{K}$ system has a bound state
just below the $K\bar{K}$ threshold or not.
According to our method of answering this question,
the condition for a non-trivial solution of the
Schr\"{o}dinger equation to
exist for the total energy of the whole system
below the lowest threshold is the one  given by eq.(\ref{detqzero}).
The the ${\sf Q}$ matrix mentioned there is actually a
complicated function of
the parameters of the formalism fitted above, except $\bar{k}$, and
the total energy of the whole system.
$\bar{k}$ (see eq.(\ref{f}))
is the phenomenological parameter of our model of gluonic effects.
Our numerical calculations, including calculation of the
determinant of the $\sf Q$ matrix, were  done for three values of
$\bar{k}$ in turn.
The value $\bar{k}=0$
corresponds to a two-body potential model hamiltonian.
On the other hand,
$\bar{k}=1/2$ ${\rm fm}^{-2}$ is emerging from the lattice
gauge theory
calculations  \cite{migp93} for rectangular configurations of
quark positions.
For other configurations, indications~  \cite{migps92}
are that the spatial
decrease of gluonic topologies overlap may be slower,
and thus we have also used an intermediate value
($\bar{k}=1/6\ {\rm fm}^{-2}$). Which of
these, if any, would simulate the
"experimental" (lattice gauge theory
calculations based) behaviour of
the gluonic overlap is yet to be seen.

Our numerical
calculations showed that for any value of  $\bar{k}$,
the above condition for the existence of a bound state (see
eq.(\ref{detqzero}))
is {\em not} satisfied
for any value of energy below the $K\bar{K}$ threshold.
This is the situation in
the isoscalar as
well as in the isovector sector, whereas in the latter case
all connections to
the $\pi\eta$ channel are neglected as those would not affect
the answer to the  main question being discussed here.
On the other hand, using a closely related model
Weinstein and
Isgur   \cite{wn90}  get $K\bar{K}$ bound states in both the
isoscalar and the isovector sectors, and conclude
that the two scalar meson resonances $f_0(975)$ and
$a_0(980)$ can be explained as loosely bound $K\bar{K}$ states.
Their model corresponds, in some approximation
(see the first paragraph of the next section),
to ours in the limit $\bar{k}=0$.
Therefore it is interesting to see if, in the corresponding
limit, we can get their results by varying our parameters.
Our calculations  showed that, in the $\bar{k}=0$ limit, we need
to multiply our total couplings of the two channels
by a factor of 2.715 before we can get bound states
in both the isoscalar and
isovector sectors. Alternatively, we can get
these bound states
by multiplying {\em only} the hyperfine couplings by
a factor of 1.926.
We get bound states without hyperfine interaction
as well, but for
that an increase by a factor of 3.06 in the remaining
couplings, with $\bar{k}=0$,
would be needed. This is one of the indications in our work that
the hyperfine coupling is the main interaction arising through quark
exchange, and that the
hyperfine and other couplings arising through confinement etc. have
opposite signs.

Our modification
to the two-body
potential model proposed in this paper (equivalent to using
non-zero values of $\bar{k}$) implies a decrease in
the $K\bar{K}$
coupling.  That  means that we need to increase the couplings
even more
so as to get binding. The factors so needed for the different
values of
$\bar{k}$ are reported in table (\ref{binding}), along with the
corresponding energy values for the resulting binding.

\begin{table}
\begin{tabular}{|c|l|c|c|l|r|}
\hline  &
\multicolumn{3}{c|}{hyperfine coupling}
& \multicolumn{1}{c|}{increase}
& \multicolumn{1}{c|}{increase} \\
\cline{2-4} $\bar{k}$ & \multicolumn{1}{c|}{increase to } &
\multicolumn{1}{c|}{B.E.
in the } & \multicolumn{1}{c|}{B.E. in the } &
\multicolumn{1}{c|}{in the total } &
\multicolumn{1}{c|}{without } \\ in ${\rm fm}^{-2}$ &
 \multicolumn{1}{c|}{get binding} & \multicolumn{1}{c|}{isovector }
& \multicolumn{1}{c|}{isoscalar }
& \multicolumn{1}{c|}{coupling to} &
\multicolumn{1}{c|}{hyperfine to} \\ & &\multicolumn{1}{c|}{channel}
&\multicolumn{1}{c|}{channel} & \multicolumn{1}{c|}{get binding} &
\multicolumn{1}{c|}{get binding} \\
\hline 0  & \hspace{.5cm} 1.926  & $<$1 MeV &
 35 MeV & \hspace{.5cm} 2.715 &  3.06  \hspace{.5cm}
\\ \hline 1/6  & \hspace{.5cm} 3.00 &
 $<$1 MeV & 53 MeV & \hspace{.5cm} 4.26 &  6.57  \hspace{.5cm}
\\ \hline 1/2 & \hspace{.5cm} 6.43 & $<$1 MeV &   87 MeV
& \hspace{.5cm} 8.89 &  19.13  \hspace{.5cm}
\\ \hline
\end{tabular}
\caption{Increases in the couplings necessary to get binding.}
\label{binding}
\end{table}

In addition we report below
(in tables (\ref{phaseb}) to (\ref{phasee}))
the $K\bar{K}$ phase shifts, in the elastic as well as
in the inelastic region,
for the hyperfine interaction increased by the above mentioned
factor of 1.926, for the different values of $\bar{k}$.
With that increase in the coupling
we get bound states of the whole $K\bar{K}$ system for $\bar{k}=0$
in both the isoscalar and
the isovector sectors, and what is
explored here is just
the effect of our proposed modification to the four-body
potential. The numerical
procedure to get these phase shifts was based on
eqs.(\ref{eps1d1}) and
(\ref{eps2d2}). Each of them is a complex equation and
hence can be solved for the two quantities $\epsilon_k$
(the inelasticity factor)
and $\delta_k$ (the phase shift), with $k=1$ or 2,
for each value of energy.

Also mentioned are  the
values of the inelastic phase shifts for the incoming
waves in the other channel
i.e. $\eta\eta$ in the isoscalar channel and
$\pi\eta'$ in the isovector one. Moreover, the values of
all of these phase shifts in the absence of hyperfine interaction
(without any increase of coupling)
are reported. However, it must be emphasized that
because of the
various approximations which we have used in this  work,
it would be improper to take these numbers as
precise results
of our model. One of the  indications of this inaccuracy is the
violation of unitarity resulting from our separable approximation to
the actual non-separable terms in the integrands appearing in our
coupled equations (\ref{cempkkc}) and (\ref{cempeec}).
As mentioned in the paragraph following eq.(\ref{approx}),
this approximation affects badly our results for the
phase shifts, although not our
conclusions regarding $K\bar{K}$ binding.
For the elastic region this
unitarity violation manifests itself in the reported (see
tables (\ref{phaseb}) to (\ref{phasee}))
deviation from unity  of the inelasticity factor $\epsilon_1$.

\begin{table}
\begin{tabular}{|rc|r|r|r|r|r|r|}
\hline & &
\multicolumn{3}{c|}{full coupling} & \multicolumn{3}{c|}{without
hyperfine} \\
\cline{3-8}
\multicolumn{2}{|c|}{$E_c$} & \multicolumn{1}{c|}{$\delta_1$} &
\multicolumn{1}{c|}{$\epsilon_1$} &
\multicolumn{1}{c|}{$\delta_2$}
& \multicolumn{1}{c|}{$\delta_1$}
& \multicolumn{1}{c|}{$\epsilon_1$} &
\multicolumn{1}{c|}{$\delta_2$} \\ \multicolumn{2}{|c|}{(total cm
energy)} & \multicolumn{1}{c|}{(degs)}
& & \multicolumn{1}{c|}{(degs)}
& \multicolumn{1}{c|}{(degs)} & & \multicolumn{1}{c|}{(degs)} \\
\hline \hspace{.25in} 997.0 & \hspace{-.4 in} MeV  & 102.09 & 0.729 &
& 1.11 & 1.00 &
\\ \hline \hspace{.25in}
1042.0 & \hspace{-.4 in} MeV  & 91.49 & 0.719 &
& 3.05 & 1.00 &
\\ \hline
\hspace{.25in} 1092.0 & \hspace{-.4 in} MeV  & 96.56 & 0.726 &
& 4.05 & 1.00 &
\\ \hline \hspace{.25in}
1095.5 & \hspace{-.4 in} MeV  & 99.01 & 0.728 &
& 4.25 & 1.00 &
\\ \hline \hspace{.25in}
1142.0 & \hspace{-.4 in} MeV  & 91.74 &  & 155.94
& 3.94 &  & 1.35
\\ \hline
\hspace{.25in} 1192.0 & \hspace{-.4 in} MeV  & 83.41 &  & 147.26
& 3.39 &  & 1.46
\\ \hline
\end{tabular}
\caption{The isovector sector phase shifts for $\bar{k}=0$.}
\label{phaseb}
\end{table}

\begin{table}
\begin{tabular}{|rc|r|r|r|r|r|r|}
\hline & &
\multicolumn{3}{c|}{full coupling} & \multicolumn{3}{c|}{without
hyperfine} \\
\cline{3-8}
\multicolumn{2}{|c|}{$E_c$}
& \multicolumn{1}{c|}{$\delta_1$}
& \multicolumn{1}{c|}{$\epsilon_1$} &
\multicolumn{1}{c|}{$\delta_2$}
& \multicolumn{1}{c|}{$\delta_1$}
& \multicolumn{1}{c|}{$\epsilon_1$} &
\multicolumn{1}{c|}{$\delta_2$} \\ \multicolumn{2}{|c|}{(total cm
energy)} & \multicolumn{1}{c|}{(degs)}
& & \multicolumn{1}{c|}{(degs)}
& \multicolumn{1}{c|}{(degs)} & & \multicolumn{1}{c|}{(degs)} \\
\hline \hspace{.25in} 997.0 & \hspace{-.4 in} MeV  & 157.18 & 0.953 &
& 2.29 & 1.00 &
\\ \hline
\hspace{.25in} 1042.0 & \hspace{-.4 in} MeV  & 125.92 & 0.804 &
& 6.56 & 1.00 &
\\ \hline
\hspace{.25in} 1092.0 & \hspace{-.4 in} MeV  & 118.85 & 0.779 &
& 9.61 & 1.00 &
\\ \hline
\hspace{.25in} 1097.5 & \hspace{-.4 in} MeV  & 119.76 & 0.781 &
& 10.73 & 1.00 &
\\ \hline
\hspace{.25in} 1142.0 & \hspace{-.4 in} MeV  & 112.10 &  & 142.78
& 9.26 &  & 4.38
\\ \hline
\hspace{.25in} 1192.0 & \hspace{-.4 in} MeV  & 105.26 &  & 127.53
& 7.33 &  & 4.45
\\ \hline
\end{tabular}
\caption{The isoscalar sector phase shifts for $\bar{k}=0$.}
\label{otherc}
\end{table}

\begin{table}
\begin{tabular}{|rc|r|r|r|r|r|r|} \hline & &
\multicolumn{3}{c|}{full coupling} & \multicolumn{3}{c|}{without
hyperfine} \\
\cline{3-8} \multicolumn{2}{|c|}{$E_c$}
& \multicolumn{1}{c|}{$\delta_1$}
& \multicolumn{1}{c|}{$\epsilon_1$} &
\multicolumn{1}{c|}{$\delta_2$}
& \multicolumn{1}{c|}{$\delta_1$}
& \multicolumn{1}{c|}{$\epsilon_1$} &
\multicolumn{1}{c|}{$\delta_2$} \\ \multicolumn{2}{|c|}{(total cm
energy)} & \multicolumn{1}{c|}{(degs)}
& & \multicolumn{1}{c|}{(degs)}
& \multicolumn{1}{c|}{(degs)} & & \multicolumn{1}{c|}{(degs)} \\
\hline \hspace{.25in} 997.0 & \hspace{-.4 in} MeV  & 5.13 & 0.997 &
& 0.21 & 1.00 &
\\ \hline \hspace{.25in} 1042.0
& \hspace{-.4 in} MeV  & 17.10 & 0.972 &
& 0.59 & 1.00 &
\\ \hline \hspace{.25in} 1092.0
& \hspace{-.4 in} MeV  & 30.15 & 0.919 &
& 0.82 & 1.00 &
\\ \hline
\hspace{.25in} 1095.5 & \hspace{-.4 in} MeV  & 33.03 & 0.905 &
& 0.86 & 1.00 &
\\ \hline
\hspace{.25in} 1142.0 & \hspace{-.4 in} MeV  & 30.84 &  & 179.46
& 0.83 &  & 0.27
\\ \hline
\hspace{.25in} 1192.0 & \hspace{-.4 in} MeV  & 27.19 &  & 177.86
& 0.74 &  & 0.30
\\ \hline
\end{tabular}
\caption{The
isovector sector phase shifts for $\bar{k}=1/6\ {\rm fm}^{-2}$.}
\end{table}

\begin{table}
\begin{tabular}{|rc|r|r|r|r|r|r|} \hline & &
\multicolumn{3}{c|}{full coupling} & \multicolumn{3}{c|}{without
hyperfine} \\
\cline{3-8}
\multicolumn{2}{|c|}{$E_c$}
& \multicolumn{1}{c|}{$\delta_1$}
& \multicolumn{1}{c|}{$\epsilon_1$} &
\multicolumn{1}{c|}{$\delta_2$} &
\multicolumn{1}{c|}{$\delta_1$} & \multicolumn{1}{c|}{$\epsilon_1$} &
\multicolumn{1}{c|}{$\delta_2$} \\ \multicolumn{2}{|c|}{(total cm
energy)} & \multicolumn{1}{c|}{(degs)} & &
\multicolumn{1}{c|}{(degs)} & \multicolumn{1}{c|}{(degs)}
& & \multicolumn{1}{c|}{(degs)} \\
\hline \hspace{.25in} 997.0 & \hspace{-.4 in} MeV  & 17.21 & 0.972 &
& 0.40 & 1.00 &
\\ \hline \hspace{.25in} 1042.0 &
\hspace{-.4 in} MeV  & 50.39 & 0.821 &
& 1.18 & 1.00 &
\\ \hline
\hspace{.25in} 1092.0 & \hspace{-.4 in} MeV  & 79.08 & 0.725 &
& 1.79 & 1.00 &
\\ \hline
\hspace{.25in} 1097.5 & \hspace{-.4 in} MeV  & 87.37 & 0.715 &
& 1.98 & 1.00 &
\\ \hline
\hspace{.25in} 1142.0 & \hspace{-.4 in} MeV  & 86.34 &  & 151.52
& 1.86 &  & 0.90
\\ \hline
\hspace{.25in} 1192.0 & \hspace{-.4 in} MeV  & 78.69 &  & 140.20
& 1.56 &  & 0.96
\\ \hline
\end{tabular}
\caption{The isoscalar
sector phase shifts for $\bar{k}=1/6\ {\rm fm}^{-2}$.}
\end{table}
\vspace{ .4 in}

\begin{table}
\begin{tabular}{|rc|r|r|r|r|r|r|} \hline & &
\multicolumn{3}{c|}{full coupling} & \multicolumn{3}{c|}{without
hyperfine} \\
\cline{3-8} \multicolumn{2}{|c|}{$E_c$} &
\multicolumn{1}{c|}{$\delta_1$} & \multicolumn{1}{c|}{$\epsilon_1$} &
\multicolumn{1}{c|}{$\delta_2$} &
\multicolumn{1}{c|}{$\delta_1$} & \multicolumn{1}{c|}{$\epsilon_1$} &
\multicolumn{1}{c|}{$\delta_2$} \\ \multicolumn{2}{|c|}{(total cm
energy)} & \multicolumn{1}{c|}{(degs)} & &
\multicolumn{1}{c|}{(degs)} & \multicolumn{1}{c|}{(degs)} & &
\multicolumn{1}{c|}{(degs)} \\
\hline \hspace{.25in} 997.0 & \hspace{-.4 in} MeV  & 0.58 & 1.000 &
& 0.02 & 1.00 &
\\ \hline \hspace{.25in} 1042.0 & \hspace{-.4 in} MeV  & 1.96 & 1.000
& & 0.06 & 1.00 &
\\ \hline \hspace{.25in} 1092.0 & \hspace{-.4 in} MeV
& 3.38 & 0.999 & & 0.09 & 1.00 &
\\ \hline \hspace{.25in} 1095.5 & \hspace{-.4 in} MeV  & 3.62
& 0.999 & & 0.10 & 1.00 &
\\ \hline \hspace{.25in} 1142.0 & \hspace{-.4 in} MeV
& 4.15 &  & 1.25 & 0.10 &  & 0.03
\\ \hline \hspace{.25in} 1192.0 &
\hspace{-.4 in} MeV  & 4.25 &  & 1.51 & 0.09 &  & 0.04
\\ \hline
\end{tabular}
\caption{The isovector sector phase shifts
for $\bar{k}=1/2\ {\rm fm}^{-2}$.}
\end{table}

\begin{table}
\begin{tabular}{|rc|r|r|r|r|r|r|} \hline & &
\multicolumn{3}{c|}{full coupling} & \multicolumn{3}{c|}{without
hyperfine} \\
\cline{3-8} \multicolumn{2}{|c|}{$E_c$} &
\multicolumn{1}{c|}{$\delta_1$} & \multicolumn{1}{c|}{$\epsilon_1$} &
\multicolumn{1}{c|}{$\delta_2$} & \multicolumn{1}{c|}{$\delta_1$} &
\multicolumn{1}{c|}{$\epsilon_1$} &
\multicolumn{1}{c|}{$\delta_2$} \\ \multicolumn{2}{|c|}{(total cm
energy)} & \multicolumn{1}{c|}{(degs)} & &
\multicolumn{1}{c|}{(degs)} & \multicolumn{1}{c|}{(degs)} & &
\multicolumn{1}{c|}{(degs)} \\
\hline \hspace{.25in} 997.0 & \hspace{-.4 in} MeV  & 1.17 & 1.000 &
& 0.04 & 1.00 &
\\ \hline \hspace{.25in} 1042.0 & \hspace{-.4 in} MeV  & 4.18
& 0.998 & & 0.12 & 1.00 &
\\ \hline \hspace{.25in} 1092.0 & \hspace{-.4 in} MeV
& 8.01 & 0.994 & & 0.19 & 1.00 &
\\ \hline \hspace{.25in} 1097.5 &
\hspace{-.4 in} MeV  & 9.22 & 0.991 & & 0.22 & 1.00 &
\\ \hline \hspace{.25in} 1142.0 & \hspace{-.4 in} MeV
& 9.28 &  & 3.83 & 0.21 &  & 0.10
\\ \hline \hspace{.25in} 1192.0 & \hspace{-.4 in} MeV
& 8.36 &  & 4.39 & 0.19 &  & 0.12
\\ \hline
\end{tabular}
\caption{The isoscalar sector phase shifts for
$\bar{k}=1/2\ {\rm fm}^{-2}$.}
\label{phasee}
\end{table}

\section{Conclusions}
\setcounter{equation}{0}
In this paper a formalism has been developed to
deal with meson-meson systems
ihaving  their dynamics  resulting through
quark exchange effects, and applied,
as a first application to a realistic case, to $K\bar{K}$ systems.
Here we had to increase our resulting coupling by some numerical
factor  before we could get a bound state of the whole
system, even in the two-body potential limit.
The variational calculations based on the two-body potential reported
in ~\cite{wn90} claim to get bound states of
the whole $K\bar{K}$ system,
concluding that the two scalar meson resonances $f_0(975)$ and
$a_0(980)$ can be explained as loosely bound $K\bar{K}$ states in the
isoscalar and the isovector sectors, respectively.
Our detailed model of meson-meson dynamics, even in the two-body
potential model limit, is different to theirs, mainly
because of our  restricted  (i.e. only in the $2S$ diagonal term)
incorporation of annihilation effects.
This neglect of the annihilation effects may have
appreciably decreased the $K\bar{K}$
binding arising through our model. This is
expected as the annihilation part of the hamiltonian, incorporating
the process $K\bar{K}\leftrightarrow\pi\pi$, was   \cite{privt}
mainly responsible for the
$K\bar{K}$ binding in the calculations reported in   \cite{wn90}.
Moreover, we might be underestimating the hyperfine interaction by
treating this interaction partially as a perturbation, although
in our work as well the hyperfine
coupling turned out to be the main interaction
arising through the quark exchange mechanism.
It is difficult to say
more about this problem unless a more refined treatment
of the hyperfine interaction, along with the annihilation effects,
is carried out. On the other hand,
the fitting of the model parameters in   \cite{wn90}
includes adjusting the ranges and normalization of
their effective meson-meson
potentials in an {\em ad-hoc} way, and it is
not clear how that affects their results.

Leaving these issues to some future work,
we looked  for any possible change in one of our parameters so as to
get $K\bar{K}$ bound states in the limit where our model would
roughly correspond to that used in
\cite{wn90} (i.e. for $\bar{k}=0$),
and then determined, with the changed parameter,
the effects of going
beyond that limit i.e.  of using our theoretically
improved {\em four-body} potential
rather than the naive two-body potential.
This means the  use of  non-zero values of $\bar{k}$ in our
terminology. This investigation showed the same trend as
observed in the spin independent case reported in ~  \cite{masu91}:
increasing $\bar{k}$, i.e. decreasing  the gluonic
states overlap, results in a significantly
weaker meson-meson
interaction.
This means that {\em if} we get a $K\bar{K}$ bound state in the
two-body potential model limit, we do not necessarily
get one with our QCD-inspired
refinement of the $q^2\bar{q}^2$ potential.

Much improvement in the calculations can be made by going beyond the
approximations we have used,
giving quantitatively more precise results. But
even without this being carried out, this work clearly indicates
that the theoretical refinement of the four-body potential results
in an appreciable decrease in
a major part  of the meson-meson interaction---enough
to cast doubt on any result based on a naive
two-body potential model.

\begin{center}
\section*{Acknowledgements}
\end{center}

I would like to express my gratitude to
J.Paton for a useful collaboration as a supervisor
during my stay in Oxford, U.K., and for help in writing this paper.
Moreover I extend my thanks
to A.M.Green for many enlightening discussions,
a fruitful collaboration,
especially during my stays in Helsinki, and for
carefully reading the typescript.
I thank the Oxford Students Scholarship Committee,
Oriel College, Oxford,
and the University of Helsinki
for their financial assistance during different
stages of this work.
I would also like to thank J.Weinstein for a helpful discussion.

\hspace{.5in}

\section*{Appendix A: The spin basis}
\renewcommand{\theequation}{A.\arabic{equation}}
\setcounter{equation}{0}
In the main part of this paper, we use the spin
basis given through the eqs.(\ref{sp1}),(\ref{sp2}) and (\ref{sp3}).
The notation used in these  equations is that of the
Appendix D of ~  \cite{wn90}. In this notation, an orthonormal spin
basis is $|S_{12}S_{\bar{3}
\bar{4}}\rangle_s$ and $|{\bf A}_{12}\cdot{\bf
A}_{\bar{3}\bar{4}}
\rangle_s$, defined through eqs.(D3)  and (D4) there.
Our remaining spin
base states are given in terms of these by eqs.(D5-D8) of
the same. These equations give easily the overlaps of our
spin base states,
written as $_s\langle kI|lJ\rangle_s$ in the results
mentioned in Appendix B, as  elements of the overlap matrix
\begin{equation}
\label{amatn}
A=\left(\begin{array}{cccccc} 1 & 0 & \sqrt{1/4} & -\sqrt{3/4} &
\sqrt{1/4} & \sqrt{3/4} \\ & 1 & -\sqrt{3/4} & -\sqrt{1/4} &
\sqrt{3/4} & -\sqrt{1/4} \\ & & 1 & 0 & -\sqrt{1/4} & \sqrt{3/4} \\ &
& & 1 & -\sqrt{3/4} &
-\sqrt{1/4} \\ & & &  & 1 & 0 \\ {\rm symmetric}
& & & & & 1 \end{array} \right),
\end{equation}
in the basis $|1S\rangle_s, |1T\rangle_s, \cdots , |3T\rangle_s$.

For the matrix
elements of the ${\bf S}_i\cdot{\bf S}_j$ operators, for
different values of the indices $i$ and $j$, in our spin basis,
we also used the results expressed
through eqs.(D9-D11) of the Appendix D of ~  \cite{wn90}.
Some of the results obtained in this way are reported here:
\begin{eqnarray}
_s\langle 1S|\left[\begin{array}{c} {\bf S}_1\cdot{\bf S}_2 \\
{\bf S}_1\cdot{\bf S}_{\bar{3}}  \\ {\bf S}_1\cdot{\bf S}_{\bar{4}}
 \\ {\bf S}_2\cdot{\bf S}_{\bar{3}}
\\ {\bf S}_2\cdot{\bf S}_{\bar{4}} \\
{\bf S}_{\bar{3}}\cdot{\bf S}_{\bar{4}} \end{array} \right]
|1S\rangle_s  & =  &
_s\langle P_{1\bar{3}}P_{2\bar{4}}|
\left[\begin{array}{c} {\bf S}_1\cdot{\bf S}_2 \\
{\bf S}_1\cdot{\bf S}_{\bar{3}}  \\ {\bf S}_1\cdot{\bf S}_{\bar{4}}
 \\ {\bf S}_2\cdot{\bf S}_{\bar{3}}
\\ {\bf S}_2\cdot{\bf S}_{\bar{4}} \\
{\bf S}_{\bar{3}}\cdot{\bf S}_{\bar{4}} \end{array} \right]
|P_{1\bar{3}}P_{2\bar{4}}\rangle_s  = \left[\begin{array}{c}
0 \\ -\frac{3}{4} \\ 0 \\ 0 \\
-\frac{3}{4} \\ 0 \end{array}\right] \hspace{.5
in}  \\
_s\langle 1S|\left[\begin{array}{c} {\bf S}_1\cdot{\bf S}_2 \\
{\bf S}_1\cdot{\bf S}_{\bar{3}}  \\ {\bf S}_1\cdot{\bf S}_{\bar{4}}
 \\ {\bf S}_2\cdot{\bf S}_{\bar{3}}
\\ {\bf S}_2\cdot{\bf S}_{\bar{4}} \\
{\bf S}_{\bar{3}}\cdot{\bf S}_{\bar{4}} \end{array} \right]
|1T\rangle_s  & =  &
_s\langle 1T|\left[\begin{array}{c} {\bf S}_1\cdot{\bf S}_2 \\
{\bf S}_1\cdot{\bf S}_{\bar{3}}  \\ {\bf S}_1\cdot{\bf S}_{\bar{4}}
 \\ {\bf S}_2\cdot{\bf S}_{\bar{3}}
\\ {\bf S}_2\cdot{\bf S}_{\bar{4}} \\
{\bf S}_{\bar{3}}\cdot{\bf S}_{\bar{4}} \end{array} \right]
|1S\rangle_s   =  \left[\begin{array}{c}
-\sqrt{\frac{3}{16}} \\ 0 \\ +\sqrt{\frac{3}{16}}
\\  +\sqrt{\frac{3}{16}} \\
0 \\ -\sqrt{\frac{3}{16}} \end{array}\right]    \\
_s\langle 1T|\left[\begin{array}{c} {\bf S}_1\cdot{\bf S}_2 \\
{\bf S}_1\cdot{\bf S}_{\bar{3}}  \\ {\bf S}_1\cdot{\bf S}_{\bar{4}}
 \\ {\bf S}_2\cdot{\bf S}_{\bar{3}}
\\ {\bf S}_2\cdot{\bf S}_{\bar{4}} \\
{\bf S}_{\bar{3}}\cdot{\bf S}_{\bar{4}} \end{array} \right]
|1T\rangle_s  & =  &
_s\langle {\bf V}_{1\bar{3}}\cdot{\bf V}_{2\bar{4}}|
\left[\begin{array}{c} {\bf S}_1\cdot{\bf S}_2 \\
{\bf S}_1\cdot{\bf S}_{\bar{3}}  \\ {\bf S}_1\cdot{\bf S}_{\bar{4}}
 \\ {\bf S}_2\cdot{\bf S}_{\bar{3}}
\\ {\bf S}_2\cdot{\bf S}_{\bar{4}} \\
{\bf S}_{\bar{3}}\cdot{\bf S}_{\bar{4}} \end{array} \right]
|{\bf V}_{1\bar{3}}\cdot{\bf V}_{2\bar{4}}\rangle_s
= \left[\begin{array}{c}
-\frac{1}{2} \\ +\frac{1}{4} \\ -\frac{1}{2} \\ -\frac{1}{2}
\\ +\frac{1}{4} \\ -\frac{1}{2} \end{array}\right] \ \ \ \ \ \ \\
_s\langle 1S|\left[\begin{array}{c} {\bf S}_1\cdot{\bf S}_2 \\
{\bf S}_1\cdot{\bf S}_{\bar{3}}  \\ {\bf S}_1\cdot{\bf S}_{\bar{4}}
 \\ {\bf S}_2\cdot{\bf S}_{\bar{3}}
\\ {\bf S}_2\cdot{\bf S}_{\bar{4}} \\
{\bf S}_{\bar{3}}\cdot{\bf S}_{\bar{4}} \end{array} \right]
|2S\rangle_s  & =  &
_s\langle 2S|\left[\begin{array}{c} {\bf S}_1\cdot{\bf S}_2 \\
{\bf S}_1\cdot{\bf S}_{\bar{3}}
\\ {\bf S}_1\cdot{\bf S}_{\bar{4}}
 \\ {\bf S}_2\cdot{\bf S}_{\bar{3}}
\\ {\bf S}_2\cdot{\bf S}_{\bar{4}} \\
{\bf S}_{\bar{3}}\cdot{\bf S}_{\bar{4}} \end{array} \right]
|1S\rangle_s  =  \left[\begin{array}{c}
+\frac{3}{8} \\ -\frac{3}{8} \\ -\frac{3}{8} \\
-\frac{3}{8} \\ -\frac{3}{8} \\ +\frac{3}{8} \end{array}\right]  \\
_s\langle 2S|\left[\begin{array}{c} {\bf S}_1\cdot{\bf S}_2 \\
{\bf S}_1\cdot{\bf S}_{\bar{3}}  \\ {\bf S}_1\cdot{\bf S}_{\bar{4}}
 \\ {\bf S}_2\cdot{\bf S}_{\bar{3}}
\\ {\bf S}_2\cdot{\bf S}_{\bar{4}} \\
{\bf S}_{\bar{3}}\cdot{\bf S}_{\bar{4}} \end{array} \right]
|2S\rangle_s  & =  &
_s\langle P_{1\bar{4}}P_{2\bar{3}}|
\left[\begin{array}{c} {\bf S}_1\cdot{\bf S}_2 \\
{\bf S}_1\cdot{\bf S}_{\bar{3}}  \\ {\bf S}_1\cdot{\bf S}_{\bar{4}}
 \\ {\bf S}_2\cdot{\bf S}_{\bar{3}}
\\ {\bf S}_2\cdot{\bf S}_{\bar{4}} \\
{\bf S}_{\bar{3}}\cdot{\bf S}_{\bar{4}} \end{array} \right]
|P_{1\bar{4}}P_{2\bar{3}}\rangle_s  = \left[\begin{array}{c}
0 \\ 0 \\ -\frac{3}{4} \\ -\frac{3}{4} \\ 0 \\ 0 \end{array}\right]
\end{eqnarray}

\section*{Appendix B: The kernels
of the integro-differential equations}
\setcounter{equation}{0}
\renewcommand{\theequation}{B.\arabic{equation}}

The results for (in general non-local)
kernels appearing  in eq.(\ref{master}),
calculated using the
procedure outlined in the two
paragraphs following this equation, are:

\begin{equation}
\label{gcalnr}
{\cal N}_{kI,kJ}=\delta_{IJ}\delta({\bf R}_k-{\bf R}'_k),
\ \ \ \ \ \ \ \ \
{\rm for \ all \ values \ of} \ k\ I\ {\rm and} \ J.
\end{equation}

\begin{equation}
\label{gcalkdr}
{\cal K}_{kI,kJ}=\delta_{IJ}\delta({\bf R}_k-{\bf
R}'_k)\left[\frac{3}{4}[\omega_{k1}
+\omega_{k2}]-\frac{f_k}{2m}\underline{\nabla}_{{\bf
R}_k}^2\right],
\end{equation}
for all values of $k,I\ {\rm and} \ J$, except those
corresponding to the $2S$ diagonal term. Here
\begin{equation}
\label{omegas}
\omega_{k1}=\frac{g_k}{2md^2_{k1}} \ \ \ \ \ {\rm and} \ \ \ \ \
\omega_{k2}=\frac{h_k}{2md^2_{k2}},
\ \ \ \ \ \ \ \ \ \ \ \ {\rm along \ with}
\end{equation}
\begin{eqnarray}
f_1=f_3 & = & \frac{2}{s+1} \ \ \ \ \ \ \ \ \ \ \ \
\ \ \ \ \ \ \ \ \ \ \
\ \ \ \ \ \ \ \ \ \ \ \ \ \ \nonumber \\
g_1=g_3=h_1=h_3 & = & \frac{s+1}{s} \nonumber \\
f_2 & = & \frac{s+1}{2s} \nonumber \\
g_2 & = & 2 \nonumber \\
\label{fghnum}
h_2 & = & \frac{2}{s}.
\end{eqnarray}
$d_{k1}$ and $d_{k2}$, for
$k=1,2$ or $3$, are the same ones which appear in
eq.(\ref{dsizek}).

\begin{eqnarray}
\label{calvc1}
{\cal V}^{cf}_{1I,1J} & = & \delta_{IJ}\delta({\bf R}_1-{\bf
R}_1')\left[-\frac{8}{3}\bar{C}-4C[d_{11}^2+d_{12}^2]\right],
\ \ \  {\rm for} \ I,J=S \ {\rm or} \  T. \ {\rm Similarly,}
\ \ \ \ \ \ \ \ \ \\
{\cal V}^{cf}_{2I,2J} & = & \delta_{IJ}\delta({\bf R}_2-{\bf
R}_2')\left[-\frac{8}{3}\bar{C}-4C[d_{21}^2+d_{22}^2]\right] \\
{\cal V}_{3I,3J}^{cf} & = & \delta_{IJ}\delta({\bf R}_3-{\bf
R}'_3)\times \nonumber \\
 & & \left\{-\frac{8}{3}\bar{C}-\frac{4}{3}C{\bf
R}_3^2-6Cd'^2-2Cd'^2\left(\frac{s-1}{s+1}\right)^2 \right.
\nonumber \\ & & +\frac{5}{2}\frac{D}
{(1+4\bar{k}d'^2)^{3/2}}\frac{[s+1]^3}
{[8\bar{k}d'^2(s^2+1)+(s+1)^2]^{3/2}}\times
\nonumber \\
 & & \exp\left[4\bar{k}(s-1)^2{\bf
R}_3^2\left(\frac{4\bar{k}d'^2}
{8\bar{k}d'^2(s^2+1)+(s+1)^2}-\frac{1}{(s-1)^2}\right)\right]
\nonumber \\
 & & -\frac{5}{2}\frac{D}
{(1+8\bar{k}d'^2)^{3/2}}\frac{[s+1]^3}{[16\bar{k}d'^2(s^2+1)
+(s+1)^2]^{3/2}}\times
\nonumber \\
 & & \left.\exp\left[8\bar{k}(s-1)^2{\bf
R}_3^2\left(\frac{8\bar{k}d'^2}{16\bar{k}d'^2(s^2+1)+
(s+1)^2}-\frac{1}{(s-1)^2}\right)\right]\right\}.\
\ \ \ \ \ \ \ \
\end{eqnarray}
(See the discussion before eq.(\ref{drat}) for $d'$).

\begin{eqnarray}
\label{calvh1ss}
{\cal V}_{1S,1S}^{hyp} &
= & -\frac{8\pi}{3m^2s}\delta({\bf R}_1-{\bf
R}'_1)\left[\frac{\alpha_s^{1\bar{3}}}{(2\pi
d^2_{11})^{3/2}}+\frac{\alpha_s^{2\bar{4}}}{(2\pi d^2_{12})^{3/2}}
\right] \\
\label{calvh1tt}
{\cal V}_{1T,1T}^{hyp} &
= & -\frac{1}{3}{\cal V}_{1S,1S}^{hyp}  \\
\label{calvh2tt}
{\cal V}_{2T,2T}^{hyp} &
= & \frac{1}{3}\frac{8\pi}{3m^2s}\delta({\bf R}_2-{\bf
R}'_2)\left[\frac{s\alpha_s^{1\bar{4}}}{(2\pi
d^2_{21})^{3/2}}+\frac{\alpha_s^{2\bar{3}}}{s(2\pi
d^2_{22})^{3/2}}\right] \\
{\cal V}_{3S,3S}^{hyp} &
= & -\frac{1}{2}\frac{8\pi}{3m^2s}\delta({\bf
R}_3-{\bf R}_3')\frac{1}{(2\pi
d'^2)^{3/2}}\left[\alpha_s^{12}+\alpha_s^{\bar{3}\bar{4}}\right] \\
{\cal V}^{hyp}_{3S,3T} & = & {\cal V}^{hyp}_{3T,3S} =
-\frac{1}{4\sqrt{3}}\frac{8\pi}{3m^2s}\delta({\bf R}_3-{\bf
R}_3')\frac{1}{(2\pi
d'^2)^{3/2}}\times \nonumber \\
 & & \left\{\left[\alpha_s^{1\bar{3}}
+\alpha_s^{2\bar{4}}\right]\left[\frac{(s+1)^2}
{(s^2+1)}\right]^{3/2}\exp\left[-\frac{{\bf
R}_3^2}{2d'^2}\frac{(s+1)^2}{s^2+1}\right]\right. \nonumber \\
 & & -s\alpha_s^{1\bar{4}}\left[\frac{(s+1)^2}
{2s^2}\right]^{3/2}\exp\left[-\frac{{\bf
R}_3^2}{2d'^2}\frac{(s+1)^2}{2s^2}\right] \nonumber \\
 & & \left. -\frac{\alpha_s^{2\bar{3}}}{s}
\left[\frac{(s+1)^2}{2}\right]^{3/2}\exp\left[-\frac{{\bf
R}_3^2}{2d'^2}\frac{(s+1)^2}{2}\right]\right\} \\
{\cal V}^{hyp}_{3T,3T} & =
& -\frac{1}{6}\frac{8\pi}{3m^2s}\delta({\bf R}_3-{\bf
R}_3')\frac{1}{(2\pi
d'^2)^{3/2}}\times \nonumber \\
 & & \left\{-\left[\alpha_s^{12}+
\alpha_s^{\bar{3}\bar{4}}\right]+[\alpha_s^{1\bar{3}}+
\alpha_s^{2\bar{4}}]\left[\frac{(s+1)^2}{(s^2+1)}
\right]^{3/2}\exp\left[-\frac{{\bf
R}_3^2}{2d'^2}\frac{(s+1)^2}{s^2+1}\right]\right. \nonumber \\
 & &  +s\alpha_s^{1\bar{4}}\left[\frac{(s+1)^2}{2s^2}\right]^{3/2}
\exp\left[-\frac{{\bf
R}_3^2}{2d'^2}\frac{(s+1)^2}{2s^2}\right] \nonumber \\
 & & \left. +\frac{\alpha_s^{2\bar{3}}}{s}
\left[\frac{(s+1)^2}{2}\right]^{3/2}\exp\left[-\frac{{\bf
R}_3^2}{2d'^2}\frac{(s+1)^2}{2}\right]\right\}. \\
\label{calnk2r}
{\cal N}_{kI,2J} & = & {e_0}
_s\langle kI|2J\rangle_s N^0_{k,2}\exp\left[-e_1{\bf R}_k^2-e_2{\bf
R}^2_2\right] \ \ \ \ \ \ \ \ \ \ {\rm for} \ \ \ \ \
k=1 \ \ {\rm or} \ \ 3 \\
\label{caln2lr}
{\cal N}_{2I,lJ} & = & {e_0}
_s\langle 2I|lJ\rangle_s N^0_{2,l}\exp\left[-e_1{\bf R}_l^2-e_2{\bf
R}^2_2\right]\ \ \ \ \ \ \ \ \ \ \ {\rm for} \ \ \ \ \ \
l=1 \ \ {\rm or} \ \ 3 \\
{\cal N}_{kI,lJ} & = &
 _s\langle kI|lJ\rangle_sN^0_{k,l}
\frac{(s+1)^6}{64s^3}\left[\frac{1}
{\pi d'^2(1+4\bar{k}d'^2)}\right]^{3/2}
\times \nonumber \\
 & & \exp\left\{-\left(\frac{s+1}{2}\right)^2
\left(\frac{1+8\bar{k}d'^2}{4d'^2}\right)
\left[\frac{s^2+1}{s^2}\left({\bf
R}_k^2+{\bf R}_l^2\right)+2\frac{s^2-1}{s^2}{\bf R}_k\cdot{\bf
R}_l\right]\right\} \nonumber \\
\label{calnkl}
   & & {\rm for}\ \ \ \ \ k,l=1 \ \ {\rm or} \ \ 3,
\ \ \ \ \ \ \ \ \ \ {\rm but
\ with} \ \ \ \ \ k\neq l. \ \ \ \ \ {\rm Here}
\end{eqnarray}
\begin{eqnarray}
\label{e0}
e_0 & = & (s+1)^{9/4}s^{-15/8}2^{3/4}
\left(\pi\kappa d^2\right)^{-3/2} \\
\label{e1}
e_1 & = & \frac{1}{4d^2}
\left(\frac{s+1}{2}\right)^2
\left[\gamma-\frac{\lambda^2}{\kappa}\right] \\
\label{e2}
e_2 & = & 4\bar{k}+\frac{1}{2d^2}\sqrt{\frac{2s}{s+1}},
\ \ \ \ \ \ \ \ \ \ {\rm with}
\end{eqnarray}
\begin{eqnarray}
\label{kappa}
\kappa & = &
8\bar{k}d^2\left[\frac{s^2+1}{s^2}\right]+1+
s^{-3/2}\left[\frac{(s+1)^2}{\sqrt{2(s+1)}}+1\right]
\\
\label{lambda}
\lambda & = &
8\bar{k}d^2\left[\frac{s^2-1}{s^2}\right]+1+
s^{-3/2}\left[\frac{s^2-1}{\sqrt{2(s+1)}}-1\right]
\\
\label{gamma}
\gamma & = &
8\bar{k}d^2\left[\frac{s^2+1}{s^2}\right]+
1+s^{-3/2}\left[\frac{(s-1)^2}{\sqrt{2(s+1)}}+1\right].
\end{eqnarray}
\begin{eqnarray}
\label{calkk2r}
{\cal K}_{kI,2J} & = &
-\frac{e_0}{2m} _s\langle kI|2J\rangle_s
N^0_{k,2}\left[q_{11}{\bf R}_k^2+q_{12}{\bf
R}_2^2+q_{10}\right]
\exp\left[-e_1{\bf R}_k^2-e_2{\bf R}_2^2\right]
\nonumber \\
 & &  \ \ \ \ \ \ \ \ {\rm for}\ \ \ \ \ \ k=1 \ \ \ {\rm
or}
\ \ \ 3 \\
\label{calk2lr}
{\cal K}_{2I,lJ} & = & -\frac{e_0}{2m}
_s\langle 2I|lJ\rangle_s N^0_{2,l}
\left[q_{21}{\bf R}_l^2+q_{22}{\bf
R}_2^2+q_{20}\right]\exp\left[-e_1{\bf R}_l^2-e_2{\bf R}_2^2\right]
\nonumber \\
 & &  \ \ \ \ \ \ \ \ {\rm for}\ \ \ \ \ \ l=1 \ \ \ {\rm
or}
\ \ \ 3
\label{calkklr} \\
{\cal K}_{kI,lJ} & = & -\frac{1}{2m}
_s\langle kI|lJ\rangle_sN^0_{k,l}\frac{(s+1)^6}{64s^3}
\left[\frac{1}{\pi d'^2(1+4\bar{k}d'^2)}\right]^{3/2}
\times \nonumber \\
 & & \left\{\left(\frac{s+1}{2}\right)^4
\left[\!\!\!\!\left[\left(\frac{s-1}{s+1}\right)^2{\bf
R}_k^2\left(\frac{8(s-1)^2}{(s+1)^3}\left[\frac{4}{(s-1)^2}
+\frac{1}{s}\right]^2\left[
\frac{1+4\bar{k}d'^2}{2d'^2}\right]^2\right.\right.\right.\right.
\nonumber \\
 & & \left. +\frac{32s}{(s+1)^3}\frac{(s+1)^4}
{s^4}\left[\frac{1+8\bar{k}d'^2}{8d'^2}\right]^2\right)
\nonumber \\
 & & -2\left(\frac{s-1}{s+1}\right){\bf R}_k\cdot{\bf
R}_l\left(\frac{8(s-1)^2}{(s+1)^3}
\frac{1}{s}\left[\frac{4}{(s-1)^2}+\frac{1}{s}\right]
\left[\frac{1+4\bar{k}d'^2}{2d'^2}\right]^2\right.
\nonumber \\
 & & \left. +\frac{32s}{(s+1)^3}
\frac{(s+1)^2}{s^2}\left[\frac{1+
8\bar{k}d'^2}{8d'^2}\right]\frac{1}{s^2}\left[
\frac{(s+1)^2+(1+s^2)8\bar{k}d'^2}{8d'^2}\right]\right)
\nonumber \\
 & &  \left.\left.+{\bf
R}_l^2
\left(\frac{8(s-1)^2}{(s+1)^3}\frac{1}{s^2}
\left[\frac{1+4\bar{k}d'^2}{2d'^2}\right]^2 +
\frac{32s}{(s+1)^3}\frac{1}{s^4}
\left[\frac{(s+1)^2+(1+s^2)8\bar{k}d'^2}{8d'^2}\right]^2\right)
\right]\!\!\!\!\right]
\nonumber \\
 & & -6
\left(\frac{s+1}{4}\right)^2
\left[\frac{8(s-1)^2}{(s+1)^3}\frac{4}{(s-1)^2}
\left(\frac{1+4\bar{k}d'^2}{2d'^2}\right)\right.
\nonumber \\
 & & \left.
\left. +\frac{32s}{(s+1)^3}\frac{1}{s^2}
\left(\frac{(s+1)^2+(1+s^2)8\bar{k}d'^2}{8d'^2}\right)\right]-3
\left(\frac{s+1}{2s}\right)
\left[\frac{1+4\bar{k}d'^2}{2d'^2}\right]\right\}\times
\nonumber \\
 & & \exp
\left\{-\left(\frac{s+1}{2}\right)^2
\left(\frac{1+8\bar{k}d'^2}{4d'^2}\right)
\left[\frac{s^2+1}{s^2}\left({\bf
R}_k^2+{\bf R}_l^2\right)+2\frac{s^2-1}{s^2}{\bf R}_k\cdot{\bf
R}_l\right]\right\} \nonumber \\
\label{calkkl}
   & &  \ \ \ \ \ \ \ \ {\rm for} \ \ \ \ \ \ k,l=1 \ \ \
{\rm or} \ \ \ 3, \ \ \ \ \ \ \ \ \ \  {\rm but
\ with} \ \ \ \ \ k\neq l. \ \ \ \ \ {\rm Here}
\end{eqnarray}
\begin{eqnarray}
q_{11} & = &
\left(\frac{s+1}{2}\right)^4
\left\{\frac{8(s-1)^2}{(s+1)^3}
\left[\left(\frac{s-1}{s+1}\right)
\left(\frac{8\bar{k}}{(s-1)^2}+\frac{1+\sqrt{s}}{(s-1)^2d^2}\right)
\right.\right. \nonumber \\
& & \left.-\left(\frac{\lambda}{\kappa}-\frac{s-1}{s+1}\right)
\left(\frac{2\bar{k}}{s}+\frac{1}{2d^2\sqrt{s}(1+\sqrt{s})}\right)
\right]^2 \nonumber \\
 & &
+\frac{32s}{(s+1)^3}\left[\left(\frac{s-1}{s+1}\right)
\left(\frac{2\bar{k}}{s}+\frac{1}{2d^2\sqrt{s}(1+\sqrt{s})}\right)
\right. \nonumber \\
\label{q11}
& & \left.\left.-\left(\frac{\lambda}{\kappa}-\frac{s-1}{s+1}\right)
\left(\bar{k}\frac{s^2+1}{s^2}+\frac{s^{-3/2}+1}{4d^2}\right)
\right]^2\right\} \\
\label{q12}
q_{12} & = &
4\left(\frac{s+1}{2s}\right)
\left[2\bar{k}+\frac{1}{2d^2}
\sqrt{\frac{2s}{s+1}}\right]^2 \\
q_{10} & = &
-\frac{3}{2}
\left(\frac{s+1}{2}\right)^2
\left[\frac{8(s-1)^2}{(s+1)^3}
\left(\frac{8\bar{k}}{(s-1)^2}+\frac{1+\sqrt{s}}{(s-1)^2d^2}\right)
\right. \nonumber \\
 & &
\left.+\frac{32s}{(s+1)^3}
\left(\bar{k}\frac{s^2+1}{s^2}+\frac{s^{-3/2}+1}{4d^2}\right)\right]
   \nonumber \\
 & & +\frac{3d^2}{2\kappa}(s+1)^2
\left[\frac{8(s-1)^2}{(s+1)^3}
\left(\frac{2\bar{k}}{s}+\frac{1}{2d^2\sqrt{s}(1+\sqrt{s})}\right)^2
\right. \nonumber \\
 & &
\left.+\frac{32s}{(s+1)^3}
\left(\bar{k}\frac{s^2+1}{s^2}+\frac{s^{-3/2}+1}{4d^2}\right)^2
\right] \nonumber \\
\label{q10}
 & & -6
\left(\frac{s+1}{2s}\right)
\left[2\bar{k}+
\frac{1}{2d^2}\sqrt{\frac{2s}{s+1}}\right] \\
\label{q21}
q_{21} & = & 2(s+1)^2
\left[
\left(1-\frac{\lambda}{\kappa}\right)^2
\left(\bar{k}+\frac{1}{4d^2}\right)^2+s
\left(1+\frac{\lambda}{\kappa}\right)^2
\left(\frac{\bar{k}}{s^2}+\frac{s^{-3/2}}{4d^2}\right)^2
\right] \ \ \ \ \ \ \ \ \\
\label{q22}
 q_{22} & = & 4
\left(\frac{s+1}{2s}\right)
\left[2\bar{k}+\frac{1}{2d^2}\sqrt{\frac{2s}{s+1}}\right]^2 \\
q_{20} & = & \frac{8}{(s+1)^2}
\left(\frac{s+1}{2}\right)^2\frac{24d^2}{\kappa}
\left[\left(\bar{k}+\frac{1}{4d^2}\right)^2+s
\left(\frac{\bar{k}}{s^2}+\frac{s^{-3/2}}{4d^2}\right)^2\right]
\nonumber \\
 & & -6\times\frac{8}{(s+1)^2}
\left(\frac{s+1}{2}\right)^2
\left[\left(\bar{k}+\frac{1}{4d^2}\right)+s
\left(\frac{\bar{k}}{s^2}+\frac{s^{-3/2}}{4d^2}\right)\right]
\nonumber \\
\label{q20}
 & & -6
\left(\frac{s+1}{2s}\right)
\left[2\bar{k}+
\frac{1}{2d^2}\sqrt{\frac{2s}{s+1}}\right].
\end{eqnarray}
\begin{equation}
\label{calvckiljr}
{\cal V}^{cf}_{kI,lJ}=
-\frac{8}{3}\bar{C}{\cal N}_{kI,lJ}+
_s\langle kI|lJ\rangle_s {\cal V}^{cf}_{k,l},     \ \ \ \ \ \ \ \ \ \
{\rm with}
\end{equation}
\begin{eqnarray}
\label{calvc12r}
{\cal V}^{cf}_{1,2} & = & {\cal V}^{cf}_{2,1} =  e_0[b_1{\bf R}_1^2+
b_0]\exp\left[-e_1{\bf R}_1^2-e_2{\bf
R}^2_2\right] \\
{\cal V}_{2,3}^{cf} & = & {\cal
V}_{3,2}^{cf}=\frac{2}{3\sqrt{3}}Ce_0\frac{(s+1)^2}{2s^2}
\left\{
\left[(s^2+s+1)\frac{\lambda^2}
{\kappa^2}-2(s^2-1)\frac{\lambda}{\kappa}+(s^2-s+1)\right]{\bf
R}_3^2\right.
\nonumber \\
 & &
\left. +\frac{(s^2+s+1)}{(s+1)^2}\frac{24d^2}{\kappa}\right\}\exp
\left[-e_1{\bf R}_3^2-e_2{\bf
R}_2^2\right] \\
{\cal V}^{cf}_{1,3} & = & {\cal V}^{cf}_{3,1}=
-\frac{2}{3\sqrt{3}}C\frac{(s+1)^6}{64s^3}
\left[\frac{1}{\pi
d'^2(1+4\bar{k}d'^2)}\right]^{3/2}\times \nonumber \\
& & \left\{\frac{1}{2s^2}
\left(\frac{s+1}{2}\right)^2
\left[(s+1){\bf
R}_3+(s-1){\bf R}_1\right]^2+\frac{9d'^2}{1+4\bar{k}d'^2}\right\}
\times \nonumber \\
 & & \exp
\left\{-
\left(\frac{s+1}{2}\right)^2
\left(\frac{1+8\bar{k}d'^2}{4d'^2}\right)
\left[\frac{s^2+1}{s^2}
\left({\bf
R}_1^2+{\bf R}_3^2\right)+2\frac{s^2-1}{s^2}{\bf R}_1\cdot{\bf
R}_3\right]\right\}. \nonumber \\
\end{eqnarray}
Here the new definitions used are
\begin{eqnarray}
\label{cb1}
b_1 & = &
-\frac{4}{9}C\frac{(s+1)^4}{4s^2}
\left[\frac{\lambda}{\kappa}-\frac{s-1}{s+1}\right]^2 \\
\label{b0}
b_0 & = & -\frac{8}{3}C
\left(\frac{s+1}{s}\right)^2\frac{d^2}{\kappa}.
\end{eqnarray}

\begin{eqnarray}
\label{calvh1s2sr}
{\cal V}^{hyp}_{1S,2S} & = & {\cal
V}^{hyp}_{2S,1S}=-\frac{e_0}{6}\frac{8\pi}{3m^2s}
\frac{(2\kappa)^{3/2}}{(2\pi d^2)^{3/2}}\left[L_{11}+
L_{12}+L_{13}\right]\times \nonumber \\ & & \exp
\left[-(e_1+e_1'){\bf R}_1^2-e_2{\bf R}_2^2\right] \\
{\cal V}^{hyp}_{1S,2T} & = & {\cal
V}^{hyp}_{2T,1S}=-\frac{e_0}{6\sqrt{3}}\frac{8\pi}{3m^2s}
\frac{(2\kappa)^{3/2}}{(2\pi d^2)^{3/2}}
\left[-3L_{11}+L_{12}+L_{13}\right]\times  \nonumber \\ & & \exp
\left[-(e_1+e_1'){\bf R}_1^2-e_2{\bf R}_2^2\right] \\
{\cal V}^{hyp}_{1T,2S} & = & {\cal
V}^{hyp}_{2S,1T}=-\frac{e_0}{6\sqrt{3}}\frac{8\pi}{3m^2s}
\frac{(2\kappa)^{3/2}}{(2\pi d^2)^{3/2}}\left[L_{11}-3
L_{12}+L_{13}\right]\times  \nonumber \\ & & \exp
\left[-(e_1+e_1'){\bf R}_1^2-e_2{\bf R}_2^2\right] \\
{\cal V}^{hyp}_{1T,2T} & = & {\cal
V}^{hyp}_{2T,1T}=-\frac{e_0}{18}\frac{8\pi}{3m^2s}
\frac{(2\kappa)^{3/2}}{(2\pi d^2)^{3/2}}\left[L_{11}+
L_{12}+5L_{13}\right]\times  \nonumber \\ & & \exp
\left[-(e_1+e_1'){\bf R}_1^2-e_2{\bf R}_2^2\right] \\
{\cal V}^{hyp}_{3S,2S} & = & {\cal
V}^{hyp}_{2S,3S}=-\frac{e_0}{4\sqrt{3}}\frac{8\pi}{3m^2s}
\frac{(2\kappa)^{3/2}}{(2\pi d^2)^{3/2}}\left[L_{21}+2
L_{22}+L_{23}\right]\times  \nonumber \\ & & \exp
\left[-(e_1+e_1'){\bf R}_3^2-e_2{\bf R}_2^2\right] \\
{\cal V}^{hyp}_{3S,2T} & = & {\cal
V}^{hyp}_{2T,3S}=-\frac{e_0}{12}\frac{8\pi}{3m^2s}
\frac{(2\kappa)^{3/2}}{(2\pi d^2)^{3/2}}\left[-L_{21}-2
L_{22}+3L_{23}\right]\times  \nonumber \\ & & \exp
\left[-(e_1+e_1'){\bf R}_3^2-e_2{\bf R}_2^2\right] \\
{\cal V}^{hyp}_{3T,2S} & = & {\cal
V}^{hyp}_{2S,3T}=-\frac{e_0}{12}\frac{8\pi}{3m^2s}
\frac{(2\kappa)^{3/2}}{(2\pi d^2)^{3/2}}
\left[L_{21}-6
L_{22}+L_{23}\right]\times  \nonumber \\
 & & \exp
\left[-(e_1+e_1'){\bf R}_3^2-e_2{\bf R}_2^2\right] \\
{\cal V}^{hyp}_{3T,2T} & = & {\cal
V}^{hyp}_{2T,3T}=\frac{e_0}{12\sqrt{3}}\frac{8\pi}{3m^2s}
\frac{(2\kappa)^{3/2}}{(2\pi d^2)^{3/2}}
\left[5L_{21}+2
L_{22}+L_{23}\right]\times  \nonumber \\
 & & \exp
\left[-(e_1+e_1'){\bf R}_3^2-e_2{\bf R}_2^2\right] \\
{\cal V}^{hyp}_{1S,3S} & = & {\cal
V}^{hyp}_{3S,1S}=-\frac{1}{4\sqrt{3}}\frac{8\pi}{3m^2s}
\frac{1}{(2\pi d'^2)^{3/2}}
\left(\frac{2}{\pi d'^2}\right)^{3/2}\frac{(s+1)^6}{64s^3}
\left[2L_{31}+L_{32}+L_{33}\right]
\times \nonumber \\
 & & \exp
\left\{-\left(\frac{s+1}{2}\right)^2
\left(\frac{1+8\bar{k}d'^2}{4d'^2}\right)
\left[\frac{s^2+1}{s^2}\left({\bf R}_1^2+
{\bf R}_3^2\right)+2\frac{s^2-1}{s^2}{\bf R}_1\cdot{\bf
R}_3\right]\right\} \\
{\cal V}^{hyp}_{1S,3T} & = & {\cal
V}^{hyp}_{3T,1S}=-\frac{1}{12}\frac{8\pi}{3m^2s}
\frac{1}{(2\pi d'^2)^{3/2}}
\left(\frac{2}{\pi d'^2}\right)^{3/2}\frac{(s+1)^6}{64s^3}
\left[6L_{31}-L_{32}-L_{33}\right]
\times \nonumber \\
 & & \exp
\left\{-\left(\frac{s+1}{2}\right)^2
\left(\frac{1+8\bar{k}d'^2}{4d'^2}\right)
\left[\frac{s^2+1}{s^2}\left({\bf R}_1^2+{\bf R}_3^2
\right)+2\frac{s^2-1}{s^2}{\bf R}_1\cdot{\bf R}_3
\right]\right\} \\
{\cal V}^{hyp}_{1T,3S} & = & {\cal
V}^{hyp}_{3S,1T}=-\frac{1}{12}\frac{8\pi}{3m^2s}
\frac{1}{(2\pi d'^2)^{3/2}}
\left(\frac{2}{\pi d'^2}\right)^{3/2}\frac{(s+1)^6}{64s^3}
\left[-2L_{31}-L_{32}+3L_{33}\right]
\times \nonumber \\
 & & \exp\left\{-\left(\frac{s+1}{2}\right)^2
\left(\frac{1+8\bar{k}d'^2}{4d'^2}\right)
\left[\frac{s^2+1}{s^2}
\left({\bf
R}_1^2+{\bf R}_3^2\right)+2\frac{s^2-1}{s^2}{\bf R}_1\cdot{\bf
R}_3\right]\right\} \\
{\cal V}^{hyp}_{1T,3T} & = & {\cal
V}^{hyp}_{3T,1T}=-\frac{1}{12\sqrt{3}}\frac{8\pi}{3m^2s}
\frac{1}{(2\pi d'^2)^{3/2}}
\left(\frac{2}{\pi d'^2}\right)^{3/2}\frac{(s+1)^6}{64s^3}
\left[2L_{31}+5L_{32}+L_{33}\right]
\times \nonumber \\
\label{calvh1t3tr}
 & & \exp\left\{-\left(\frac{s+1}{2}\right)^2
\left(\frac{1+8\bar{k}d'^2}{4d'^2}\right)
\left[\frac{s^2+1}{s^2}
\left({\bf
R}_1^2+{\bf R}_3^2\right)+2\frac{s^2-1}{s^2}{\bf R}_1\cdot{\bf
R}_3\right]\right\}. \ \ \ \ \ \ \ \
\end{eqnarray}
Here
\begin{eqnarray}
\label{capl11}
L_{11} & = & l_{10}\exp
\left[l_{11}{\bf
R}^2_1-l_{12}{\bf R}_2^2\right]
\left\{\alpha^{1\bar{3}}_s\exp
\left[l_{13}{\bf R}_1\cdot{\bf
R}_2\right]+\alpha^{2\bar{4}}_s\exp
\left[-l_{13}{\bf
R}_1\cdot{\bf
R}_2\right]\right\}  \ \ \ \ \ \ \ \ \ \ \ \ \\
\label{capl12}
L_{12} & = & l_{20}\exp\left[-l_{21}{\bf
R}_1^2\right]
\left\{\alpha^{1\bar{4}}_s\exp
\left[l_{22}{\bf
R}_1^2\right]+s\alpha_s^{2\bar{3}}\exp
\left[-l_{22}{\bf
R}_1^2\right]\right\} \\
\label{capl13}
L_{13} & = & l_{30}\exp
\left[-l_{31}{\bf R}_1^2-l_{32}{\bf
R}_2^2\right]
\left\{\alpha^{12}_s\exp
\left[-l_{33}{\bf R}_1\cdot{\bf
R}_2\right]+\alpha_s^{\bar{3}\bar{4}}\exp
\left[l_{33}{\bf R}_1\cdot{\bf R}_2\right]\right\} \\
L_{21} & = & l_{30}\exp
\left[-l_{31}{\bf R}_3^2-l_{32}{\bf
R}_2^2\right]
\left\{\alpha_s^{1\bar{3}}\exp
\left[-l_{33}{\bf R}_3\cdot{\bf
R}_2\right]+\alpha_s^{2\bar{4}}\exp
\left[l_{33}{\bf R}_3\cdot{\bf R}_2\right]\right\} \\
L_{22} & = & l_{20}\exp
\left[-l_{21}{\bf
R}_3^2\right]
\left\{\alpha^{1\bar{4}}_s\exp
\left[l_{22}{\bf
R}_3^2\right]+s\alpha_s^{2\bar{3}}\exp
\left[-l_{22}{\bf
R}_3^2\right]\right\} \\
L_{23} & = & l_{10}\exp
\left[l_{11}{\bf R}_3^2-l_{12}{\bf
R}_2^2\right]
\left\{\alpha^{12}_s\exp
\left[l_{13}{\bf R}_3\cdot{\bf
R}_2\right]+\alpha_s^{\bar{3}\bar{4}}\exp
\left[-l_{13}{\bf R}_3\cdot{\bf R}_2\right]\right\} \\
L_{31} & = &
\left[\alpha_s^{1\bar{3}}+\alpha_s^{2\bar{4}}\right]\exp
\left\{-\frac{(s+1)^4}{16s^2}\frac{1+4\bar{k}d'^2}{d'^2}
\left[\left(\frac{s-1}{s+1}\right){\bf
R}_1+{\bf R}_3\right]^2\right\} \\
L_{32} & = &
\left[\frac{\pi d'^2}{1+4\bar{k}d'^2}\right]^{3/2}
\left[\frac{\phantom{d^2}2\phantom{d^2}}{s+1}\right]
\left\{s\alpha_s^{1\bar{4}}\delta({\bf
R}_1+{\bf R}_3)+\alpha^{2\bar{3}}_s\delta({\bf
R}_1-{\bf R}_3)\right\} \\
L_{33} & = &
\left[\alpha_s^{12}+\alpha_s^{\bar{3}\bar{4}}\right]\exp
\left\{-\frac{(s+1)^4}{16s^2}\frac{1+4\bar{k}d'^2}{d'^2}
\left[{\bf R}_1+
\left(\frac{s-1}{s+1}\right){\bf
R}_3\right]^2\right\}, \ \ \ \ \ {\rm with}
\end{eqnarray}
\begin{eqnarray}
\label{e1p}
e_1' & = &
\frac{1}{4d^2}
\left(\frac{s+1}{2}\right)^2\frac{\lambda^2}{\kappa} \\
\label{l10}
l_{10} & = &
\left(\frac{s}{s+1}\right)^3 \\
\label{l11}
l_{11} & = & \frac{s^2-1}{16d^2}
\left[2\lambda-\kappa
\left(\frac{s-1}{s+1}\right)\right] \\
\label{l12}
l_{12} & = & \frac{\kappa}{d^2}
\left(\frac{s}{s+1}\right)^2 \\
\label{l13}
l_{13} & = & \frac{s}{2d^2}
\left[\lambda-\kappa
\left(\frac{s-1}{s+1}\right)\right] \\
\label{l20}
l_{20} & = & \frac{s}{8} \\
\label{l21}
l_{21} & = & \frac{\kappa}{4d^2}
\left(\frac{s+1}{2}\right)^2 \\
\label{l22}
l_{22} & = & \frac{\lambda}{2d^2}
\left(\frac{s+1}{2}\right)^2 \\
\label{l30}
l_{30} & = &
\left(\frac{s}{s-1}\right)^3 \\
\label{l31}
l_{31} & = & \frac{1}{4d^2}
\left(\frac{s+1}{2}\right)^2
\left(\frac{s+1}{s-1}\right)
\left[-2\lambda+\kappa
\left(\frac{s+1}{s-1}\right)\right] \\
\label{l32}
l_{32} & = & \frac{\kappa}{d^2}
\left(\frac{s}{s-1}\right)^2 \\
\label{l33}
l_{33} & = & \frac{s}{2d^2}
\left(\frac{s+1}{s-1}\right)\left[-\lambda+\kappa
\left(\frac{s+1}{s-1}\right)\right].
\end{eqnarray}

As mentioned in the text,
in the
case of the diagonal term corresponding to
the $2S$ channel, the expressions
depend upon the  physical content of this  channel. Thus we have:
\begin{equation}
\label{caln2r}
{\cal N}_{2S,2S}=\delta({\bf R}_2-{\bf R}'_2),
\end{equation}
( irrespective of the physical pseudoscalar mesons present).

\begin{eqnarray}
{\cal K}_{2S,2S}(\eta\eta) & = & \delta({\bf R}_2-{\bf
R}'_2)\times \nonumber \\
& &
\left[\frac{3}{4}[(\omega^l_{21}+\omega^l_{22})
\cos^2\theta+(\omega^s_{21}+\omega^s_{22})\sin^2\theta]\right.
\nonumber \\
\label{calkeer}
& & \left.-\frac{1}{2m}(f_2^{ll}\cos^4\theta+
2f_2^{ls}\cos^2\theta\sin^2\theta+f_2^{ss}\sin^4\theta)
\underline{\nabla}_{{\bf
R}_2}^2\right], \nonumber \\
\end{eqnarray}
\begin{eqnarray}
\label{omegas1}
{\rm with}  \ \ \ \ \ \ \ \ \
\omega^l_{21} & = & \frac{g^l_2}{2md_{21}^2}=\frac{g^l_2}{2md^2},
 \ \ \ \ \ \  \ \ \ \ \ \ \ \ \ \ \ \ \ \ \
\omega^s_{21}=\frac{g^s_2}{2m
d_{22}^2}=\frac{g^s_2}{2m{d''}^2}, \ \ \ \ \ \ \ \ \ \ \ \\
\label{omegas2}
\omega^l_{22} & = & \frac{h^l_2}{2md_{21}^2}=\frac{h^l_2}{2md^2}\
\ \ \ \ \ \ \ \ {\rm and}  \ \
\ \ \ \ \ \ \
\omega^s_{22}=\frac{h^s_2}
{2md_{22}^2}=\frac{h^s_2}{2m{d''}^2}. \ \ \ \ \ \ \ \ \ \ \
\end{eqnarray}

\begin{eqnarray}
({\rm Here} \ \ \ \ \ \ \ \ \ \
f_2^{ll} & = & 1, \ \ \ \ \ f_2^{ls}=\frac{1}{2}(1+1/s), \ \ \ \ \
f_2^{ss}=1/s \nonumber \\
g_2^l & = & h_2^l=2 \ \ \ \ \ {\rm and} \ \ \ \ \
g_2^s=h_2^s=2/s. )
\end{eqnarray}
\[{\cal K}_{2S,2S}(\pi\eta) =  \delta({\bf R}_2-{\bf
R}'_2)\times \]
\begin{equation}
\label{calkper}
\left[\frac{3}{4}[\omega^l_{21}+\omega^l_{22}
\cos^2\theta+\omega^s_{22}\sin^2\theta]-
\frac{1}{2m}(f_2^{ll}\cos^2\theta+f_2^{ls}
\sin^2\theta)\underline{\nabla}_{{\bf
R}_2}^2\right].
\end{equation}

\[{\cal K}_{2S,2S}(\pi\eta') =  \delta({\bf R}_2-{\bf
R}'_2)\times \]
\begin{equation}
\label{calkpepr}
\left[\frac{3}{4}[\omega^l_{21}+\omega^l_{22}
\sin^2\theta+\omega^s_{22}\cos^2\theta]-\frac{1}{2m}(f_2^{ll}
\sin^2\theta+f_2^{ls}\cos^2\theta)\underline{\nabla}_{{\bf
R}_2}^2\right].
\end{equation}

\begin{eqnarray}
\label{calvceer}
{\cal V}^{cf}_{2S,2S}(\eta\eta) & = & \delta({\bf R}_2-{\bf
R}_2')
\left[-\frac{8}{3}\bar{C}
-2\times4C[d_{21}^2\cos^2\theta+d_{22}^2
\sin^2\theta]\right] \ \ \ \ \  \\
\label{calvcper}
{\cal V}^{cf}_{2S,2S}(\pi\eta) & = &  \delta({\bf R}_2-{\bf
R}_2')
\left[-\frac{8}{3}\bar{C}-4C[d_{21}^2+d_{21}^2\cos^2\theta+
d_{22}^2
\sin^2\theta]\right] \\
\label{calvcpepr}
{\cal V}^{cf}_{2S,2S}(\pi\eta') & = &  \delta({\bf R}_2-{\bf
R}_2')
\left[-\frac{8}{3}\bar{C}
-4C[d_{21}^2+d_{21}^2
\sin^2\theta+d_{22}^2\cos^2\theta]\right]. \\
\label{calvheer}
{\cal V}^{hyp}_{2S,2S}(\eta\eta) &
= & -\frac{8\pi}{3m^2s}\delta({\bf R}_2-{\bf R}_2')
\left[\frac{s(\alpha_s^{1\bar{4}(l)}+
\alpha_s^{2\bar{3}(l)})\cos^2\theta}{(2\pi
d_{21}^2)^{3/2}}+\frac{(\alpha_s^{1\bar{4}(s)}+
\alpha_s^{2\bar{3}(s)})
\sin^2\theta}{s(2\pi
d_{22}^2)^{3/2}}\right] \hspace{ .4 in} \\
\label{calvhper}
{\cal V}^{hyp}_{2S,2S}(\pi\eta) &
= & -\frac{8\pi}{3m^2s}\delta({\bf
R}_2-{\bf R}_2')
\left[\frac{s\alpha_s^{1\bar{4}(l)}}
{(2\pi d_{21}^2)^{3/2}}+
\frac{s\alpha_s^{2\bar{3}(l)}\cos^2\theta}{(2\pi
d_{21}^2)^{3/2}}+\frac{\alpha_s^{2\bar{3}(s)}
\sin^2\theta}{s(2\pi
d_{22}^2)^{3/2}}\right] \\
\label{calvhpepr}
{\cal V}^{hyp}_{2S,2S}(\pi\eta') & = & -\frac{8\pi}{3m^2s}\delta({\bf
R}_2-{\bf R}_2')
\left[\frac{s\alpha_s^{1\bar{4}(l)}}
{(2\pi d_{21}^2)^{3/2}}+\frac{s\alpha_s^{2\bar{3}(l)}
\sin^2\theta}{(2\pi
d_{21}^2)^{3/2}}+\frac{\alpha_s^{2\bar{3}(s)}\cos^2\theta}{s(2\pi
d_{22}^2)^{3/2}}\right]. \ \ \ \ \ \ \ \ \ \\
\label{calvaeer}
{\cal V}^a_{2S,2S}(\eta\eta) & = &
2\times\delta({\bf R}_2-{\bf R}_2')
\left[2l\cos^2\theta-2\sqrt{2ln}\cos\theta
\sin\theta+n\sin^2\theta\right]\
\ \ \ \  \ \ \ \ \ \\
\label{calvaper}
{\cal V}^a_{2S,2S}(\pi\eta) & = &\delta({\bf R}_2-{\bf R}_2')
\left[2l\cos^2\theta-2\sqrt{2ln}\cos\theta
\sin\theta+n
\sin^2\theta\right] \\
\label{calvapepr}
{\cal V}^a_{2S,2S}(\pi\eta') & = &\delta({\bf R}_2-{\bf R}_2')
\left[2l\sin^2\theta+2\sqrt{2ln}\cos\theta
\sin\theta+n\cos^2\theta\right].
\end{eqnarray}

\section*{Appendix C: The definitions used
in the momentum space solutions}
\setcounter{equation}{0}
\renewcommand{\theequation}{C.\arabic{equation}}
The following definitions are used in
writing the momentum space solutions
(\ref{gsole11}) and (\ref{gsole12}) of the coupled equations.

\begin{eqnarray}
Q_1^{(1)} & = &
\left[-\frac{1}{2m}
\frac{q_{11}}{6}+\frac{b_1}{2}\right]F_b(p_1,e_1)+
\left[-\frac{1}{2m}\frac{q_{10}}{6}
+\frac{b_0}{2}-\frac{E'_c}{6}\right]F_a(p_1,e_1)
\nonumber \\
 & & -H\alpha^{ll}_sl_{20}F_a(p_1,e_1+e'_1+l_{21}-l_{22})
-Hs\alpha_s^{ss}l_{20}F_a(p_1,e_1+e'_1+l_{21}+l_{22})
\nonumber \\
Q^{(1)}_2 & = & -\frac{1}{2m}\frac{q_{12}}{6}F_a(p_1,e_1) \
nonumber  \\
Q^{(1)}_3 & = & -2H\alpha^{ls}_sl_{10}
F_a(p_1,e_1+e'_1-l_{11}) \nonumber \\
Q^{(1)}_4 & = &
-\frac{1}{3}H\alpha_s^{ls}
l_{10}l_{13}^2F_b(p_1,e_1+e'_1-l_{11}) \nonumber \\
Q^{(1)}_5 & = & -Hn_1\alpha^{ls}_sl_{30}F_a(p_1,
\tau_1\overline{e_1+e'_1+l_{31}+e_2+l_{32}-l_{33}}) \nonumber \\
\label{q1sdef}
Q^{(1)}_6 & = & Hn_1\alpha^{ls}_sl_{30}F_a(p_1,
\tau_1\overline{e_1+e'_1+l_{31}+e_2+l_{32}+l_{33}}) \\
Q^{(2)}_1 & = &
-\frac{1}{2m}\frac{q_{22}}{6}F_b(p_2,e_2)+
\left[-\frac{1}{2m}\frac{q_{20}}{6}+\frac{b_0}{2}
-\frac{E'_c}{6}\right]F_a(p_2,e_2)  \nonumber \\
Q^{(2)}_2 & = &
\left[-\frac{1}{2m}\frac{q_{21}}{6}+
\frac{b_1}{2}\right]F_a(p_2,e_2) \nonumber \\
Q^{(2)}_3 & = & -H\alpha^{ll}_sl_{20}F_a(p_2,e_2)  \nonumber \\
Q^{(2)}_4 & = & -Hs\alpha^{ss}_sl_{20}F_a(p_2,e_2)  \nonumber \\
Q^{(2)}_5 & = &
-2H\alpha_s^{ls}l_{10}F_a(p_2,e_2+l_{12}) \nonumber \\
Q^{(2)}_6 & = &
-\frac{1}{3}H\alpha_s^{ls}l_{10}l_{13}^2F_b(p_2,e_2+l_{12})
\nonumber \\
Q^{(2)}_7 & = & -Hn_2\alpha^{ls}_sl_{30}F_a(p_2,
\tau_2\overline{e_1+e'_1+l_{31}+e_2+l_{32}-l_{33}}) \nonumber \\
\label{q2sdef}
Q^{(2)}_8 & = & Hn_2\alpha^{ls}_sl_{30}F_a(p_2,
\tau_2\overline{e_1+e'_1+l_{31}+e_2+l_{32}+l_{33}}).
\end{eqnarray}

\begin{eqnarray}
\label{ecp}
E'_c & = & E_c+\frac{8}{3}\bar{C}-2m(s+1) \\
\label{ak}
A_k(x) & = & e_0 \int d^3{\bf R}_k \exp
\left[-x{\bf
R}_k^2\right]\chi_{kS}({\bf R}_k) \\
\label{bk}
B_k(x) & = & e_0 \int d^3{\bf R}_k\exp
\left[-x{\bf
R}_k^2\right]{\bf R}_k^2\chi_{kS}({\bf R}_k) \\
\label{chip}
\chi_{kS}({\bf p}_k) & = & \int \frac{d^3{\bf
R}_k}{(2\pi)^{3/2}}\exp
\left[i{\bf p}_k\cdot{\bf
R}_k\right]\chi_{kS}({\bf R}_k),
\end{eqnarray}
for $k=1,2$. $F_a({\bf p}_k,x)$ and  $F_b({\bf p}_k,x)$
are similar FTs of $\exp
\left[-x{\bf R}_k^2\right]$ and
${\bf R}_k^2\exp
\left[-x{\bf R}_k^2\right]$ respectively. Moreover
\begin{eqnarray}
\label{del1def}
\Delta_1(p_1) & = &
\frac{p_1^2}{2\mu_{K\bar{K}}}+M_K+M_{\bar{K}}-E_c-i\varepsilon \\
\label{del2def}
\Delta_2(p_2) & = &
\frac{p_2^2}{2\mu_{ab}}+M_a+M_{b}-E_c-i\varepsilon \\
\label{pc1def}
p_c(1) & = & \sqrt{2\mu_{K\bar{K}}(E_c-M_K-M_{\bar{K}})} \\
\label{pc2def}
p_c(2) & = & \sqrt{2\mu_{ab}(E_c-M_a-M_b)}.
\end{eqnarray}

\section*{Appendix D: The elements of the $T$ matrix}
\setcounter{equation}{0}
\renewcommand{\theequation}{D.\arabic{equation}}
Consistent with our
definition of the $T$ matrix (see eqs.(\ref{eps1d1})
and (\ref{eps2d2})),
the four elements of the
$2\times 2$ $T$ matrix are (these can be read off
from  eqs.(\ref{gsole11}) and (\ref{gsole12})):
\begin{eqnarray}
T_{1,1} & = &
2\mu_{K\bar{K}}\frac{\pi}{2}p_c(1)
\left[Q_1^{(1)}A_2(e_2)+Q_2^{(1)}
B_2(e_2)+Q_3^{(1)}A_2(e_2+l_{12})+Q_4^{(1)}
B_2(e_2+l_{12})\right. \nonumber \\
& & +Q_5^{(1)}A_2(\tau_1
\overline{e_1+e'_1+l_{31}+e_2+l_{32}-l_{33}})
\nonumber \\
\label{gt11ex}
& & \left.+Q_6^{(1)}A_2(\tau_1
\overline{e_1+e'_1+l_{31}+e_2+l_{32}+l_{33}})\right] \\
T_{2,1} & = &
2\mu_{ab}\frac{\pi}{2}p_c(1)\sqrt{\frac{v_2}{v_1}}
\left[Q_1^{(2)}A_1(e_1)+Q_2^{(2)}B_1(e_1)\right.
\nonumber \\
& & +Q_3^{(2)}A_1(e_1+e'_1+l_{21}-l_{22})+
Q_4^{(2)}A_1(e_1+e'_1+l_{21}+l_{22})
\nonumber \\
& & +Q_5^{(2)}A_1(e_1+e'_1-l_{11})+
Q_6^{(2)}B_1(e_1+e'_1-l_{11}) \nonumber \\
& & +Q_7^{(2)}A_1(\tau_2
\overline{e_1+e'_1+l_{31}+e_2+l_{32}-l_{33}})
\nonumber \\
\label{gt21ex}
& & \left.+Q_8^{(2)}A_1(\tau_2
\overline{e_1+e'_1+l_{31}+e_2+l_{32}
+l_{33}})\right] \\
T_{2,2} & = &
2\mu_{ab}\frac{\pi}{2}p_c(2)
\left[Q_1^{(2)}A_1(e_1)+Q_2^{(2)}B_1(e_1)\right.
\nonumber \\
& &
+Q_3^{(2)}A_1(e_1+e'_1+l_{21}-l_{22})+
Q_4^{(2)}A_1(e_1+e'_1+l_{21}+l_{22})
\nonumber \\
& & +Q_5^{(2)}A_1(e_1+e'_1-l_{11})+Q_6^{(2)}
B_1(e_1+e'_1-l_{11}) \nonumber \\
& & +Q_7^{(2)}A_1(\tau_2
\overline{e_1+e'_1+l_{31}+e_2+l_{32}-l_{33}})
\nonumber  \\
\label{gt22ex}
& & \left.+Q_8^{(2)}A_1(\tau_2
\overline{e_1+e'_1+l_{31}
+e_2+l_{32}+l_{33}})\right] \\
T_{1,2} & = &
2\mu_{K\bar{K}}\frac{\pi}{2}p_c(1)\sqrt{\frac{v_1}{v_2}}
\left[Q_1^{(1)}A_2(e_2)+Q_2^{(1)}B_2(e_2)+
Q_3^{(1)}A_2(e_2+l_{12})+Q_4^{(1)}
B_2(e_2+l_{12})\right. \nonumber \\
& & +Q_5^{(1)}A_2(\tau_1
\overline{e_1+e'_1+l_{31}+e_2+l_{32}-l_{33}})
\nonumber \\
\label{gt12ex}
& & \left.+Q_6^{(1)}A_2(\tau_1
\overline{e_1+e'_1+l_{31}+e_2+l_{32}+l_{33}})
\right],
\end{eqnarray}
with $p_1$ and $p_2$ in $Q_1^{(1)}\cdots Q_8^{(2)}$
(see eqs.(\ref{q1sdef})
and  (\ref{q2sdef})) replaced by
$p_c(1)$ and $p_c(2)$ respectively.
It was checked numerically that $T_{1,2}=T_{2,1}$, fulfilling the
requirement of ``reciprocity'' in an  inelastic scattering (see
p.528 of   \cite{blatt}).

\end{document}